\documentclass[11pt]{article}
\usepackage{amsmath, amssymb}
\usepackage{graphicx}
\usepackage{enumerate}
\usepackage{natbib}

\usepackage{enumitem}
\usepackage{mathrsfs, booktabs, amsthm, amsfonts, bbold}
\usepackage{xcolor}
\usepackage{hyperref}

\def\T{{\mathrm{T}}}

\newcommand{\bbR}{\mathbb{R}}

\protected\def\[#1\]{\begin{equation}\begin{aligned}#1\end{aligned}\end{equation}}
\protected\def\(#1\){\begin{equation*}\begin{aligned}#1\end{aligned}\end{equation*}}

\newcommand{\be}{\begin{equation*}\begin{aligned} }
\newcommand{\ee}{\end{aligned}\end{equation*} }

\newcommand{\bel}{\begin{equation}\begin{aligned} }
\newcommand{\eel}{\end{aligned}\end{equation} }

\newtheorem{theorem}{Theorem}

\newtheorem{lemma}{Lemma}

\theoremstyle{remark}
\newtheorem{remark}{Remark}
\newtheorem{assumption}{Assumption}

\theoremstyle{definition}

\newtheorem{example}{Example}

\renewenvironment{proof}[1][\proofname]{\bigskip {\noindent\bfseries Proof #1.}}{\qed}

\usepackage[labelsep=period]{caption}
\usepackage{subcaption}
\usepackage{float}
\usepackage{overpic}

\begin{document}

\title{Gradient-bridged Posterior:\\
   Bayesian Inference for Models with Implicit Functions}

\author{Cheng Zeng
	\thanks{Joint first authors}
	\thanks{Department of Statistics, University of Florida, U.S.A. {czeng1@ufl.edu}}\quad\quad
	Yaozhi Yang
	\footnotemark[1]
    \thanks{Department of Statistics, University of Florida, U.S.A.
    {yangyaozhi@ufl.edu}}\quad\quad
	Jason Xu
	\thanks{Department of Biostatistics, University of California Los Angeles, U.S.A.
	{jqxu@g.ucla.edu}}\quad\quad
	Leo L Duan
	\thanks{Department of Statistics, University of Florida, U.S.A. \href{email:li.duan@ufl.edu}{li.duan@ufl.edu}}
	\thanks{Corresponding author.}
}

\maketitle

\begin{abstract}
Many statistical problems include model parameters that are defined as the solutions to optimization sub-problems. These include classical approaches such as profile likelihood as well as  modern applications involving flow networks or Procrustes distances. In such cases, the likelihood of the data involves an implicit function, often complicating inferential procedures and entailing prohibitive computational cost. In this article, we propose an intuitive and tractable posterior inference approach for this setting. We introduce a class of continuous models that handle implicit function values using the first-order optimality of the sub-problems. Specifically, we apply a shrinkage kernel to the gradient norm, which retains a probabilistic interpretation within a generative model. This can be understood as a generalization of the Gibbs posterior framework to newly enable concentration around partial minimizers in a subset of the parameters. We show that this method, termed the gradient-bridged posterior, is amenable to efficient posterior computation, and enjoys theoretical guarantees, establishing a Bernstein--von Mises theorem for asymptotic normality. The advantages of our approach  are highlighted on a synthetic flow network experiment and an application to data integration using Procrustes distances.
\end{abstract}

\noindent Keywords: First order optimality; Interior point method; Linear programming; Data integration; Constraint relaxation.

\section{Introduction}

\label{sec:intro}
In modern statistical applications, it is often desirable to posit that some parameters arise as a solution to an optimization sub-problem. Here, a sub-problem refers to a separate loss function, which may or may not be the same as the negative-log-likelihood that characterizes the stochastic component of a model. For example, in Procrustes data analysis widely used for examining shape or microarray data, the Procrustes distance is defined as the metric between two sets of points after applying optimal rotation and reflection; thus, the rotation and reflection parameters are solutions to a Procrustes alignment sub-problem \citep{dryden2016statistical,goodall1991procrustes}. Similarly, in flow network analysis, as encountered in transportation or communication systems, it is common to model the observed flow as a noisy version of an optimal flow. The latter sub-problem is routinely defined as a linear program \citep[chapter 12]{bazaraa2011linear}.

The need to conduct statistical inference under these settings, especially under finite, moderate sample sizes, makes it appealing to consider a Bayesian approach. To be concrete, we may use the following likelihood,
\[\label{eq:bridged_posterior}
L(y; \beta,z_{\beta}) \propto g(y; \beta, \hat z_{\beta}),\qquad \text{subject to } \hat z_{\beta}={\arg\min}_{z} h(\beta,z;y),
\]
with an appropriate prior $\beta\sim \pi_0$. Under the Bayesian paradigm, estimation uncertainty is characterized through the posterior $\Pi(\beta\mid y)$. 
When the sub-problem loss $h=- \log g$, \eqref{eq:bridged_posterior} becomes a profile likelihood \citep{murphy2000profile, maclaren2018profile}; there have been many studies that justify this in the Bayesian setting \citep{lee2005profile, cheng2008higher, cheng2009penalized}. More generally, \cite{zeng2024bridged} coined the term \textit{bridged posterior} for \eqref{eq:bridged_posterior}, as it bridges a divide between Bayesian and optimization methodologies. The bridged posterior can be broadly viewed as a generalization of the projected posterior, for which $h$ is a projection loss $h(\beta,z;y)= \text{dist}(z,\beta)$ for some appropriate distance and $z$ in some constrained space \citep{sen2018constrained,chakraborty2022rates,lee2023post}.

Conceptually, the posterior associated with \eqref{eq:bridged_posterior} is an equality-constrained posterior \citep{gelfand1992bayesian}. In principle, one can fit the bridged posterior computationally by simply running an optimization algorithm nested within a Markov chain Monte Carlo sampler.
However, the lack of explicit forms for $\hat z_{\beta}$ and hence $L(y;\beta,\hat z_\beta)$ limit interpretability. These also translate into obstacles to inference, complicating procedures that are typically straightforward from a Bayesian perspective such as model-based imputation for missing values. In some cases, these difficulties can be reconciled: for instance, for some pairs $g$ and $h$ involving convex conjugate functions, \cite{polson2016mixtures} showed that an equivalent mixture representation may exist where $g(y;\beta,\hat z_\beta)\propto \int \tilde g (y;\beta, \gamma) \pi(\gamma) \textup{d}\gamma $, with both $g$ and $\pi$ tractable. Our contributions seek a more comprehensive treatment to  fill this gap when such specific representations are unavailable.

When it comes to constructing a Bayesian model motivated by a loss function, the Gibbs posterior \citep{jiang2008gibbs} is one of the most popular choices. By exponentiating a negative loss function  $\exp\{-\lambda\tilde h(\theta;y)\}$ with some $\lambda>0$, mimicking the usual role of the log-likelihood in standard Bayesian inference, one defines a distribution that concentrates around the minimizer $\hat\theta=\arg\min_\theta\tilde h(\theta;y)$. Application of Gibbs posteriors, also falling under the generalized Bayes label, covers a wide range of models \citep{bissiri2016general, martin2022direct, bhattacharya2022gibbs, rigon2023generalized, west2024perspectives}.

Unfortunately, the Gibbs posterior cannot accommodate the implicit function setting of \eqref{eq:bridged_posterior}. This is due to the two-way dependence of $h(\beta,z;y)$ on $z$ and $\beta$, given $y$. A Gibbs posterior approach would force both $\beta$ and $z$ to concentrate near a point $(\hat \beta,\hat z)={\arg\min}_{(\beta,z)} h(\beta,z;y)$. Instead, we desire to properly account for the variability of $\beta$, as \eqref{eq:bridged_posterior} is based on a \emph{partial} minimization of $h(\beta,z;y)$ over $z$ only. To illustrate, consider a simple loss $h(\beta,z;y)= \tau_1\|z-\beta\|^2_2 + \tau_2\|y-z\|^2_2/2$, with $\beta,z,y\in \mathbb{R}^d$, and $\tau_1,\tau_2>0$. We can immediately see that partial minimization of $h(\beta,z;y)$ over $z$ yields $z=(\tau_1\beta+\tau_2 y)/(\tau_1+\tau_2)$, a convex combination of $y$ and $\beta$. However, the Gibbs posterior $\exp\{ - \lambda h(\beta,z;y)\}$ instead concentrates at $(\hat \beta,\hat z): \hat\beta=\hat z=y$ for large values of $\lambda$.

Motivated to develop a conceptually simpler and computationally more efficient alternative to \eqref{eq:bridged_posterior}, we propose a continuous posterior distribution that concentrates around the partial minimizer of $h$ over $z$ for any $\beta$, for the class of models where loss in the sub-problem is differentiable. We propose to exploit the equivalence of $\hat z_{\beta}={\arg\min}_{z} h(\beta,z;y)$ and $\nabla_z h(\beta,z;y) \mid_{z=\hat z_\beta}=0$ over $z$ in an open set. This is known as the first-order optimality condition in optimization, and motivates a shrinkage kernel we impose on the gradient norm to induce a continuous distribution surrounding the partial minimizer $z \approx \hat z_\beta$. 
In doing so, the likelihood becomes tractable up to an unimportant normalizing constant, enabling straightforward inference. Elaborating on a simple core idea, we carefully establish statistical guarantees via a Bernstein-von Mises theorem, and detail several nuanced cases including its applicability in non-open domains and dual formulations. The method is amenable to efficient posterior computation via gradient-based samplers, and its merits are showcased empirically via synthetic and real data case studies.

\section{Method}

\subsection{Gradient-bridged posteriors via first-order optimality}
Throughout this article, we focus on differentiable $h( \beta, z; y)$. In this section, we first consider the case where the feasible region of $h$ is an open and convex set $\mathcal Z \subseteq \mathbb{R}^d$, and the minimizer $\arg\min_z h( \beta, z; y)$ exists and satisfies the first-order optimality condition:\[\label{eq:foc}
\hat z_\beta = \arg\min h(\beta,z;y) \text{ if and only if } \nabla_z h(\beta,\hat z_\beta;y)=0.
\]
Since $h$ also depends on $\beta$ and $y$, we assume the above condition holds almost surely with respect to the distribution of $\beta$ for all observed data $y$.

We propose the following posterior distribution:
\[\label{eq:gradient-bridged}
&\Pi(\beta, z\mid y) \propto  L(y,z;  \beta ) \pi_0(\beta), \\
& L(y,z ; \beta) \propto g(y ; \beta, z) \exp\bigl\{ - \lambda \|\nabla_z h( \beta, z; y)\|^2_2 \bigr\},
\]
where $g$ is a kernel, and the hyper-parameter $\lambda>0$  controls the degree that $z$ concentrates around $\hat z_\beta$. We see that \eqref{eq:gradient-bridged} promotes the gradient stationarity condition in \eqref{eq:foc} via regularization, and can be viewed as a relaxation of the constraint that $\nabla_z h = 0$ \citep{duan2020}. We follow \cite{presman2023} and choose a Gaussian-type kernel incorporating the squared norm $\|\nabla_z h( \beta, z; y)\|^2_2$, a choice which will facilitate gradient-based posterior computation downstream. In particular, its derivative with respect to $(\beta,z)$ is $2 [\nabla^2_{z\beta} h( \beta, z; y), \nabla^2_{zz} h( \beta, z; y)] ^\T [\nabla_z h( \beta, z; y)]$, which remains numerically stable when $\|\nabla_z h( \beta, z; y)\|\approx 0$.

Conceptually, $L(y;\beta) =\int L(y,z ; \beta) \textup{d} z$ can be interpreted as the marginal likelihood. From this lens, $z$ is a latent variable defined under the likelihood and does not require a prior.
We refer to \eqref{eq:gradient-bridged} as the \textit{gradient-bridged posterior}, and $\exp\{ - \lambda \|\nabla_z h( \beta, z; y)\|^2_2 \}$ as the shrinkage kernel.

At a large but finite $\lambda$, we effectively create a concentration of $z$ near its conditional optimal, that is, a ball $\mathbb B_\beta(\epsilon)=\{z:\|\nabla_z h(\beta, z;y)\|_2\le \epsilon\}$ has a high posterior probability for some $\epsilon>0$ given $\beta$. Indeed, as $\lambda\to\infty$, we have  $z\to\hat z_\beta$ due to $\|\nabla_z h(\beta, z;y)\|_2\ \to 0$, recovering the bridged posterior \eqref{eq:bridged_posterior}. The case of convex $h$ lends  more intuition: by the first-order characterization of convexity, we know for any $z\in \mathbb B_\beta(\epsilon)$, the optimality gap is bounded above by 
\(
h(\beta,z;y)- h(\beta,z_\beta;y)\le (z-\hat z_\beta)^{\rm T} \nabla_z h(\beta,z;y)\le \epsilon\|z-\hat z_\beta\|_2.
\)
Further, if $h$ is $\mu$-strongly convex with Lipschitz gradient, then $h(\beta,z;y)- h(\beta,\hat z_\beta;y) \le \epsilon^2/(2 \mu)$. We will develop an interesting characterization on the set $\mathbb B_\beta(\epsilon)$ in Section 2.3. To be illustrative for now, we first examine a simple example related to the normal means problem.

\begin{example}\label{example:linearreg}
	The normal means problem considers $y_i \sim \text{No}(z_i, \tau),\;z_i\sim \text{No}(0,\beta)$ independently for $i=1,\ldots,n$, with parameter $\beta>0$ and with $\tau>0$ a known constant. Due to the large number of $z_i$, one may think about reducing its variability via minimizing a loss $h(\beta,z;y)= \|z-y\|^2_2/(2\tau) + \|z\|^2/(2\beta)$, with partial derivative $\nabla_z h(\beta,z;y) = (z-y)/\tau+ z/\beta$. In this toy example, the minimization problem is tractable with minimizer $\hat z_\beta =\{1-\tau/(\tau + \beta)\}y$, which coincides with the James--Stein shrinkage estimator if $(\tau+\beta)$ is assigned the empirical estimate $\|y\|_2^2/(n-2)$. Ignoring that we know the closed form solution for now, it is interesting to examine the gradient-bridged posterior under some finite $\lambda$:
	\(
	\Pi(z\mid y,\beta) & \propto \exp\bigg( - \frac{\|z-y\|^2_2}{2\tau} - \frac{\|z\|^2}{2\beta}  -\lambda \biggl \| \frac{z-y}{\tau}+  \frac{z}{\beta}\biggr\|_2^2 \bigg),\\
	& \propto \exp\bigg\{ - \frac{\|z-\hat z_\beta\|^2_2}{2(1/\tau + 1/\beta)^{-1}}  -\lambda  \frac{\|z-\hat z_\beta\|^2}{(1/\tau + 1/\beta)^{-2}}\bigg\}.
	\)

We see that as the degree of relaxation decreases when $\lambda$ increases from $0$ to $\infty$, the distribution of $z$ gradually changes from $\text{No}[ \hat z_\beta, (1/\tau+1/\beta)^{-1} ]$ to a point mass at the solution $\hat z_\beta$.
\end{example}

\begin{remark}
We clarify that though first-order optimality is often used to characterize the solution of convex problems, the scope of our method applies beyond convex $h$. First, 
the condition \eqref{eq:foc} applies broadly to some non-convex functions. 
For example, the mode of the inverse gamma distribution is associated with the loss $f(z)=\alpha\log z+ \beta/z$, which is pseudo-convex but non-convex. Second, for non-convex problems where \eqref{eq:foc} does not hold, we can often find its Lagrange dual objective function, which is always concave and whose maximizer determines $\hat z_\beta$ (under suitable conditions). For those problems, we can form \eqref{eq:gradient-bridged} using the dual form of $h$. Further details are discussed in Section \ref{sec:non-convex}.
\end{remark}

An immediate advantage of the gradient-bridged posterior \eqref{eq:gradient-bridged} over its exact counterpart under  \eqref{eq:bridged_posterior} lies in its associated computation. By relaxing the equality constraint, we avoid having to solve a constrained optimization problem within a Bayesian procedure. Instead, one can directly apply canonical Markov chain Monte Carlo algorithms (such as random-walk Metropolis or Hamiltonian Monte Carlo), as well as alternatives such as variational inference, to estimate the posterior. Details on posterior sampling are discussed in Section 3.

\subsection{Adjustment of loss function for boundary optimum}\label{sec:boundary}
We now extend the applicable range of \eqref{eq:gradient-bridged} to allow for cases where the feasible region $\mathcal Z$ may not be open.
Many problems have a convex but non-open $\mathcal Z$, in which 
the optimal solution may fall on the boundary.
For example, in the linear programming problem where $\mathcal Z=\{z:Mz\le c\}$ for some matrix $M$ and vector $c$, one often has optimal $M^{\rm T}_j z=c_j$ for one or more indices $j$. In such cases, with $\tilde h$ the loss function, the first order optimality condition for $\hat z$  becomes $(z-\hat z)^{\rm T}\nabla_z \tilde h(\beta,\hat z;y)\ge 0$ for any $z\in \mathcal Z$, rendering the shrinkage kernel in \eqref{eq:gradient-bridged} unhelpful.

We address these cases with an  interior point approach. Suppose we have the original loss function $\tilde h$ with boundary optimum. We adjust $\tilde h$ with log-barrier function 
\(
h(\beta,z;y)= \tilde  h(\beta, z;y) - \frac{1}{t}\sum_{j=1}^m\log[r_j(z)],
\)
where the $r_j(z)$ correspond to all inequality constraints $r_j(z)\ge 0$ defining $\mathcal Z$, with hyperparameter $t>0$.  Since $-\log\{r_j(z)\}\to \infty$ as $r_j(z)\to 0$, the minimizer $\arg\min_{z} h(\beta,z;y)$
must shift from the boundary of $\mathcal Z$ into its interior. Therefore, $\hat z_\beta =\arg\min h(\beta,z;y)$ is equivalent to
\[\label{eq:foc_int}
\nabla_z  h(\beta,z;y)= \nabla_z  \tilde h(\beta, z;y) - \frac{1}{t}\sum_{j=1}^m \frac{\nabla_z r_j(z) }{r_j(z)}=0,
\]
enabling us to use $\exp\{ - \lambda \|\nabla_z  h( \beta, z; y)\|^2_2 \}$ for use in the gradient-bridged posterior.
Typically, one chooses a relatively large $t$, so that $\arg\min \tilde h(\beta,z;y)\approx \arg\min  h(\beta,z;y)$; in this article, we use $t=1000$.

We now present our first example of a Bayesian model defined with an implicit function that lacks a closed-form solution. This problem is motivated by applications in flow network data.
\begin{example}\label{example:flow}
	A flow network consists of weighted values $z_{ij}$ on a directed network $G=(V,E)$ with $V$ the set of nodes, and $E$ the set of uni-directed edges $(i\to j)$. The problem of finding the maximum flow possible from a designated source node 
	$s$ to a sink node $t$, subject to edge capacities is a well-studied linear program:
	\(
	\text{maximize} \quad & f(z)=\sum_{j:\,(s\to j) \in E} z_{sj} \\
	\text{subject to} \quad & \sum_{j:\,(i\to j) \in E} z_{ij} \, \, - \sum_{k:\,(k\to i) \in E} z_{ji} = 0, \quad \forall i \in V \setminus \{s, t\},  \\
	& 0 \le z_{ij} \le \beta_{ij}, \quad \forall (i\to j) \in E.
	\)
	In the above, the objective function  is the total amount of flow entering the network. The first equality constraint is the flow conservation: no flow is lost or created at any node except at the source and sink. The second inequality incorporates the \textit{edge capacities} $\beta_{ij}$ together with non-negativity of flows. 
	It is clear that the solution $\arg\max_z f(z)$ depends on the value of $\beta$, and is denoted by $\hat z_\beta$. For networks with more than one source,  an auxiliary  {\em super source} can be introduced to cast the problem equivalently; a similar treatment can be done for networks with more than one sink.
	
	In fields such as transportation science, it is often reasonable to assume that the flow network has nearly reached its maximum during a particular time period (such as rush hour). On the other hand, the observed flows $y_{ij}$ are realistically a noisy measurement of $z_{ij}$, and the true capacities $\beta_{ij}$ may differ from a \textit{designed capacity} $c_{ij}$. For instance, if $c_{ij}$ are lanes, congestion or closure may cause the number of usable lanes $\beta_{ij}<c_{ij}$, so that $\beta_{ij}$ may be better treated as unknown. Accounting for these considerations leads to a statistical problem; we can consider the likelihood
	\begin{equation}\label{eq:flowlikelihood}
		L(y,z;\beta) \propto \exp\bigg\{ -\frac{ \sum_{(i\to j)\in E}(y_{ij}-z_{ij})^2}{2\sigma^2_y}
		-\frac{ \sum_{(i\to j)\in E}(c_{ij}-\beta_{ij})^2}{2\sigma^2_c}
		\bigg\}
		\exp\bigl\{ - \lambda \|\nabla_z  h( \beta, z; y)\|^2_2 \bigr\},
	\end{equation}
	where it is natural to define a subproblem loss $ h( \beta, z; y)= -f(z)  - (1/t)\sum_{l=1}^m\log[r_l(z)]$ as the negative objective in addition to log-barrier terms. Here   $r_l(z)= z_{ij}$ to ensure $z_{ij}>0$, and $r_{l}(z)= \beta_{ij}-z_{ij}$ to ensure $z_{ij} < \beta_{ij}$,  for all edges $(i\to j) \in E$. To meet the equality constraint, we use a simple reparameterization: for each $i\in V\setminus\{s,t\}$, we choose one inflow edge $k^*\to i$ and parameterize $z_{k^*i}= (\sum_{j:\,(i\to j) \in E} z_{ij}) - (\sum_{k\neq k^*:\,(k\to i) \in E} z_{ki})$ as a transformation of other flows associated with $i$. We will revisit this example in our numerical study in Section 5.
\end{example}

\subsection{Implicit function manifold and its relaxation}

We now further characterize the posterior defined under the implicit function and its relaxation. In this section, we restrict our scope to the class of models where $h$ is twice continuously differentiable in $z$, and $\nabla_z h(\beta,z;y)$ is continuously differentiable in $\beta$. Within this class, the function $\nabla_z h(\beta,z;y)$ can be nicely characterized by the implicit function theorem, so that $\{(\beta,\hat z_\beta): \beta\in \Theta\}$
is a smooth manifold that we will call the \textit{implicit function manifold}.

We first briefly review the implicit function theorem using our notation. 
Let $(\beta_0, z_0)$ be a point satisfying
$\nabla_z h(\beta_0,z_0; y) = 0$,
and assume that the Hessian matrix of $h$ with respect to $z$,
$\nabla^2_{zz} h(\beta_0,z_0; y)$
is invertible. Then, by the implicit function theorem, there exists a neighborhood of $\beta_0$ in which we can define a continuously differentiable function
$$
\hat z_\beta :=  \zeta(\beta): \nabla_z h(\beta, \zeta(\beta); y) = 0.
$$
Two useful results follow. First, the function $\zeta(\beta)$ and hence $\hat z_\beta$ is unique. Second, the change rate of $\hat z_\beta $ with respect to $\beta$ is given by
$$
\frac{\partial \hat z_\beta }{\partial \beta} = -\bigl\{\nabla^2_{zz} h(\beta, \zeta(\beta); y)\bigr\}^{-1}  \nabla^2_{z\beta} h(\beta, \zeta(\beta); y),
$$
where $\nabla^2_{z\beta} h(\beta,z; y)$ denotes the matrix of mixed partial derivatives of $h$ with respect to $z$ and $\beta$.

Now, recall the gradient-bridged posterior allows $\|\nabla_z h(\beta, z; y)\|$ to be small but non-zero, and hence the high-density posterior region of $(\beta,z)$ is a continuous relaxation from the implicit function manifold. Naturally, one may wonder how much relaxation is induced at some $\|\nabla_z h(\beta, z; y)\|\le \epsilon$. The following theorem provides a bound on the amount of relaxation.

\begin{theorem} \label{thm:distance}
	At any $\beta$ with $\nabla^2_{zz} h(\beta,z_0; y)$ invertible and $z_0=\hat z_\beta$, there exists a Hessian-based neighborhood of $z_0$:
	$$\mathbb{B}(z_0,k; \beta)= \{z:
	\|I-\nabla^2_{zz} h(\beta,z_0; y)^{-1}\nabla^2_{zz} h(\beta,z; y) \|_{\text{op}}
	\le k<1\}.$$
	Further, for any $z\in \mathbb{B}(z_0,k; \beta)$, if $\|\nabla_z h(\beta, z; y)\|\le \epsilon$, we have 
	$$\|z-z_0\|\le \frac{\epsilon}{(1-k)\lambda_{\min}\{\nabla^2_{zz} h(\beta,z_0; y)\}},$$
	where $\lambda_{\min}$ denotes the smallest eigenvalue and $\| \cdot \|_{\text{op}}$ denotes the operator norm.
\end{theorem} 
\begin{remark}
	In the above we started with a Hessian-based neighborhood of $z_0$, which should not be confused with the Euclidean neighborhood of $z_0$. This Hessian-based neighborhood can be quite large when the Hessian is slowly changing as we move away from $z_0$.
\end{remark}

Roughly speaking, the above theorem shows that $\|z- \hat z_\beta\| = \mathcal{O}\{\| \nabla_z h(\beta,z;y)\|\}$, justifying that controlling the gradient norm of $h$ governing  the implicit function manifold is an effective alternative to solving for $\hat z_\beta$.

\subsection{Suitability for canonical Bayesian inference}
 
The likelihood in the gradient-bridged posterior is designed to be tractable (up to a constant). As a result, it is convenient to apply canonical Bayesian inference procedures in addition to parameter estimation under a fixed sample size $n$, such as handling missing values or model selection via Bayes factor.

To illustrate, consider the missing data setting where $y_D = \{y_i:i\in D\}$ denotes the set of observed data, and $y_M = \{y_i:i\in M\}$ those with missing values, so that the index sets $D\cup M= (1,\ldots,n)=:[n]$. Under the gradient-bridged setup, 
the joint distribution 
$$
\Pi(y_M, \beta \mid y_D,z)  \propto \pi_0(\beta) g( \{ y_D,y_M\} ; \beta, z) \exp\bigl \{ - \lambda \|\nabla_z h ( \beta, z; \{y_D,y_M\})  \|^2_2 \bigr\},$$
is amenable to simultaneous imputation of $y_M$ and estimation of $\beta$ using the same approaches that one would take in a canonical Bayesian model, for instance via data-augmented  Markov chain Monte Carlo \citep{tanner1987calculation,van2001}. As one can imagine,  the imputation  would become considerably more complex if we replace $z$ with $\hat z_\beta$, due to a constraint resulting from the deterministic relationship between $\hat z_\beta$ and $y_D$, and in turn between $\hat z_\beta$ and $y_M$.

This is just one example of how our framework inherits straightforward inference under fixed sample size $n$. Extra caution should be exercised for inference tasks under changing $n$; prediction is one such prominent example. In particular, Bayesian prediction tasks typically assign the same form of likelihood \eqref{eq:gradient-bridged} for both sample sizes $n$ and $(n+m)$, with $m$ the number of points to be predicted. A coherent prediction rule should ensure the marginalization property  $L(y_{[n]}; \beta) = \int L(y_{[n+m]} ; \beta)\, \textup{d} (y_{n+1},\ldots,y_{n+m})$ holds, as exhibited by a sequential generative process. 

Unfortunately, we do not expect this condition to hold in complete generality for gradient-bridged posterior models. This is a common limitation to loss-based generalized Bayes approaches \citep{natarajan2024cohesion, rigon2023generalized}. Nonetheless, we can identify salient sufficient conditions that ensure the marginalization condition, such as separability of the loss over $i$. To be concrete, if (i) each $y_i$ has a corresponding latent $z_i$ so that the loss  $h( \beta, z_{[n]}; y_{[n]}) = \sum_{i=1}^{n} s(z_i,y_i; \beta)$, and (ii) the marginalization condition
\[\label{eq:marg_cond}
\int  \frac{g(y_{[n+1]};\beta,z_{[n+1]})}{g(y_{[n]};\beta,z_{[n]})} \exp\bigl\{- \lambda \|\nabla_{z_{n+1}} s(z_{n+1},y_{n+1}; \beta)\|_2^2\bigr\}\,\textup{d}(y_{n+1},z_{n+1})
\]
does not depend on $\beta$, then it follows that $L(y_{[n]},z_{[n]}; \beta)= \int L(y_{[n+1]},z_{[n+1]}; \beta)\, \textup{d}(y_{n+1},z_{n+1})$. Integrating both sides over $z_{[n]}$ reveals the marginalization property. 

\begin{remark}
	In the case that \eqref{eq:marg_cond} depends on $\beta$, which we abbreviate as some function $m(\beta)$,  one can calibrate the function $g$ using a term involving $1/m(\beta)$. To illustrate, in Example \ref{example:linearreg}, the normalizing term of $g(y_{[n]};\beta,z_{[n]})$ is  $\beta^{-n/2}$. If we instead model by replacing this with $\{m(\beta)\}^{-n}$, $m(\beta)=\tau\beta\{\tau\beta+2\lambda(\tau+\beta)\}^{-1/2}$, then \eqref{eq:marg_cond} will be equal to a constant in terms of $\beta$. 
\end{remark}

\subsection{Non-convex problems and duality}\label{sec:non-convex}

We now consider \eqref{eq:gradient-bridged} in the non-convex setting, which includes the case where the function $h$ is non-convex as well as the possibility that $\mathcal Z$ is a non-convex set. Though first-order stationarity no longer suffices for optimality, progress can be made when the loss enjoys a manageable dual form.
Recall we seek to minimize what we now refer to as the \textit{primal} objective or loss
\(
f(z) :=  h(\beta,z;y),
\)
over a feasible region $\mathcal Z=\{ z: r(z)\le 0, \;s(z)= 0 \}$.

The functions $r:\mathbb{R}^d\to \mathbb{R}^{m_1}$ and $s:\mathbb{R}^d\to \mathbb{R}^{m_2}$ may depend on $y$ and $\beta$ as well. Using Lagrange multipliers $\gamma\in \{(\gamma_1,\gamma_2): \gamma_1\in \mathbb{R}^{m_1},\gamma_{1}\ge 0$, $ \gamma_2\in \mathbb{R}^{m_2} \} := \Gamma$, we have the dual function (that we will refer to as the \textit{dual} objective):
\(
f^{\dagger}(\gamma) = \min_{z\in\mathbb{R}^d} \bigl \{ f(z) + \gamma_1^\T r(z) + \gamma_2^\T s(z)
\bigr \}.
\)
Here, the minimization of $z$ is now unconstrained over $\mathbb{R}^d$. To provide some background, for any feasible $z$, we have $\gamma_1^\T r(z)\le 0$ and $\gamma_2^\T s(z)=0$ for any $\gamma\in \Gamma$. It always holds that $f^{\dagger}(\gamma)\le \min_{z\in \mathcal Z} f(z)$: this is known as \textit{weak duality}, which applies for any primal problem regardless of whether it is convex. The inequality implies two important consequences. First, the optimal dual objective $\max_{\gamma} f^{\dagger}(\gamma)$ serves as a lower-bound for the optimal primal loss. As a result, the quantity $\{f(z)- \max_{\gamma} f^{\dagger}(\gamma)\}$ provides a practical upper-bound estimate on $\{f(z)- \min_{z\in \mathcal Z} f(z)\}$ for any value $z$. Second, if we have $f(z^*)=f^{\dagger}(\gamma)$, for some pair $z^*\in \mathcal Z$ and $\gamma \in \Gamma$, then $z^*$ must be the optimal primal solution. This latter scenario is also known as \textit{strong duality}.

Importantly, $f^\dagger(\gamma)$ is always concave, as it comprises an affine function of $\gamma$ composed with a pointwise infimum over $z$. The concavity of $f^{\dagger}(\gamma)$ means that we can find the optimal $\gamma$ via the first order optimality \eqref{eq:foc}, or approximately in the interior of $\Gamma$ via \eqref{eq:foc_int}, provided $f^\dagger$ is tractable and differentiable. Using $h^{\dagger}(\beta,\gamma; y)=f^\dagger(\gamma)$ or its log-barrier modification in Section \ref{sec:boundary}, the shrinkage kernel $\exp\bigl\{ - \lambda \|\nabla_\gamma  h^{\dagger}(\beta, \gamma; y)\|^2_2 \bigr\}$ is useful toward a relaxed solution of $\arg\max_\gamma f^\dagger (\gamma)$.

\begin{remark}
	For simplicity, we will focus on primal $f$ that enjoys strong duality $\max_{\gamma\in \Gamma} f^\dagger (\gamma)= \min_{z\in \mathcal Z} f(z)$, and hence we do not need to worry about the gap between $f(z)$ and $f^\dagger (\gamma)$.
\end{remark}

To illustrate this approach in action, we consider the  orthogonal Procrustes problem \citep{gower1975generalized}, which will also play a role in our case study. The primal problem here is non-convex, but we show how the form of the dual objective enables us to construct a shrinkage kernel using \eqref{eq:foc_int}.
\begin{example}\label{example:procrustes}
	The orthogonal Procrustes problem aims to solve
	\(
	\min_{R}\;f(R)=  \min_{R}\;\|Ry- \beta\|^2_F, \quad \text{subject to } R^\T R=I_p,
	\)
	for two matrices $y\in \mathbb{R}^{p\times m}$ and $\beta\in \mathbb{R}^{p\times m}$,  
	where $R$ can include rotations and/or reflections of each column of $y$. The primal problem is not convex due to the non-convexity of the orthonormal space, although owing to von Neumann's trace inequality, the primal solution given $\beta$ is tractable using the singular value decomposition $\beta y^{\rm T}=U\Sigma V^{\rm T} $ and $\hat R= UV^{\rm T}$.  On the other hand, when $\beta$ is not known (or contains missing values) as in a Bayesian model, simultaneous updating of $R$ and $\beta$ becomes challenging.
	
	To derive the dual form, we introduce a symmetric multiplier $\gamma\in \mathbb{R}^{p\times p}$ in the Lagrangian:
	\(
	\mathcal L(R,\gamma) = \|Ry- \beta\|^2_F+ \text{tr} \{ \gamma(R^\T R-I)\},
	\)
	which is bounded from below whenever $(\gamma+yy^\T)\succ 0$ and is minimized at  $\hat R=  \beta y^\T (\gamma+y y^\T)^{-1}$.
	The finite dual function  and its derivative are given by 
	\begin{align}
		f^{\dagger}(\gamma ) \;&=\; \text{tr}(\beta^\T\beta)-\text{tr}(\gamma)    - \text{tr}\{ y \beta^\T \beta y^\T (\gamma+y y^\T)^{-1}\} , \nonumber \\
		\nabla_\gamma f^{\dagger}(\gamma )  \;&=\; -I + (y y^\T+\gamma)^{-1}\, (y \beta^\T \beta y^\T)\,(y y^\T+\gamma)^{-1} \;=\; -I + \hat R^\T \hat R. \label{eq:dualgrad}
	\end{align}
	
	Now, from the first-order optimality condition on the dual $\nabla_\gamma f^{\dagger}(\gamma ) =0$, rearranging \eqref{eq:dualgrad} shows that the solution satisfies the constraint $\hat R^\T \hat R=I_p$. That is $\hat R$ achieves primal feasibility, and strong duality holds: $\max f^{\dagger}(\gamma ) = \min_{R:R^\T R=I} f(R)$. In practice,  we further may avoid explicit matrix inversion by parameterizing a lower-triangular matrix $W$ with positive diagonal entries, taking $(y y^\T+\gamma)^{-1}=WW^\T$ which is always positive definite. Doing so, we arrive at the shrinkage kernel for the orthogonal Procrustes problem 
	\[\label{eq:procrutes_shrinkage} 
	\exp\{-\lambda \|WW^\T (y \beta^\T \beta y^\T)WW^\T-I \|^2_F \} \;:=\; \exp\{ -\lambda \|R^\T R -I \|^2_F \}
	\]
	with primal variable $R=\beta y^\T(WW^\T)$.
	
\end{example}

\section{Hamiltonian Monte Carlo Near Implicit Function Manifold} \label{subsec:mass}
The gradient-bridged posterior is designed for efficient computing, and benefits from the large body of work on gradient-based samplers and widely adopted auto-differentiation algorithms. We first review the Hamiltonian Monte Carlo algorithm, focusing on how the choice of the mass matrix can improve its efficiency. For simplicity, we abbreviate the parameter by $\theta= (z,\beta)$.

Hamiltonian Monte Carlo augments the parameters $\theta$ with an auxiliary momentum variable $p$, and makes use of Hamiltonian dynamics in the augmented space to generate informed proposals to new states. Denote the joint density  
\( \Pi(\theta, p \mid y) \propto \Pi(\theta \mid y) \, \exp \biggl(-\frac{1}{2} p^\T M^{-1} p \biggr),\)
where $M$ is a positive-definite mass matrix. A new proposal is generated by first sampling the momentum $p \sim \text{No}(0, M)$, and then simulating Hamiltonian dynamics, typically using a numerical integrator. The leapfrog algorithm is a prevalent such approach: given the current state $(\theta_t, p_t)$, we perform $L$ steps of the following, with step-size $\epsilon$, to propose $(\theta^*, p^*)$: 

\begin{enumerate}[leftmargin=2em]
	\item Half-step for momentum:
	$ p(t + {\epsilon}/{2}) = p_t + ({\epsilon}/{2}) \nabla_\theta \log \Pi(\theta_t \mid y).$
	\item Full-step for position:
	$ \theta(t + \epsilon) = \theta_t + \epsilon\, M^{-1} p(t + {\epsilon}/{2}) $.
	\item Another half-step for momentum:
	update the momentum using the new position:
	$ p(t + \epsilon) = p(t + \frac{\epsilon}{2}) + ({\epsilon}/{2} )\nabla_\theta \log \Pi(\theta(t+\epsilon) \mid y) $.
\end{enumerate}

Since the leapfrog integrator is reversible and volume preserving \citep{neal2011mcmc}, the acceptance ratio calculation is straightforward: $\theta^*$ is accepted with probability
\( \alpha\bigl\{(\theta_t, p_t), (\theta^*, p^*)\bigr\} = \min \Bigl[1, \exp\bigl\{-H(\theta^*, p^*) + H(\theta_t, p_t)\bigr\} \Bigr],\)
where the Hamiltonian is given by
$ H(\theta, p) = -\log \Pi(\theta \mid y) + \frac{1}{2} p^\T M^{-1} p. $ By simulating the Hamiltonian dynamics, the algorithm is able to propose distant moves with high acceptance rates, especially when the mass matrix $M$ is chosen judiciously to match the local geometry of the target distribution. 

An extension of Hamiltonian Monte Carlo, known as the No-U-Turn Sampler, adaptively determines the number of leapfrog steps to avoid inefficient trajectories that double back on themselves, thereby eliminating the need to pre-specify $L$ \citep{hoffman2014no}. We adopt the No-U-Turn Sampler in this article. We now focus on the choice of $M$ specifically for the gradient-bridged posterior. Combining the discrete moves in a leapfrog step, we have
\(
\theta(t + \epsilon) & = \theta_t + M^{-1} \big \{\epsilon\,  p_t + \,({\epsilon}^2/{2})  \nabla_\theta \log \Pi(\theta_t \mid y)\big \}.
\)
For large $\lambda$, we have small $\|\nabla_z h( \beta, z; y)\|^2_2 \approx 0$ with high  probability. This suggests that $\theta(t)$ lies near the manifold, and thus after a small time $\epsilon$, $\theta(t+\epsilon)$ should tend to stay near the manifold as well due to conservation of energy in the exact Hamiltonian dynamics. Since the leapfrog algorithm provides only an approximation to the exact Hamiltonian dynamics, it is sensible to choose $M$ so to avoid an abrupt change of $\|\nabla_z h( \beta, z; y)\|^2_2$.

To this end, we consider linearizing the gradient. Let $(\Delta \beta,\Delta z)=\theta(t+\epsilon) - \theta(t)$: 
then the gradient at $\theta(t+\epsilon)$ can be approximated by
$$\nabla_z h(\beta+\Delta \beta,z+\Delta z;y) \approx \nabla_z h(\beta,z;y)+ G_{\beta,z}^{\rm T}(\Delta \beta ,\Delta z),$$
where $G_{\beta,z} =[ \nabla_{z\beta}^2 h(\beta,z;y), \nabla_{zz}^2 h(\beta,z;y)]$  is the submatrix of the Hessian containing the partials with respect to both $(\beta,z)$. Since having $G_{\beta,z}^{\rm T}(\Delta \beta ,\Delta z)= 0$ would be useful for reducing changes in $\nabla_z h(\beta+\Delta \beta,z+\Delta z;y)$, this leads to a natural choice of $M^{-1}$ as a null-space projection $I - G_{\beta,z}(G_{\beta,z}^\T G_{\beta,z})^{-1} G^\T_{\beta,z}$. 

On the other hand, the leapfrog integrator would lose reversibility if $M^{-1}$ varies with the parameters, and the momentum distribution would become degenerate if $M^{-1}$ is rank-deficient.
With these practical considerations in mind, we use
\[\label{eq:precondition}
M^{-1}= (1+\tau) I - \hat G_{\beta,z}(\hat G_{\beta,z}^\T \hat G_{\beta,z})^{-1} \hat G^\T_{\beta,z},
\]
where $\tau>0$ is a small constant chosen to make $M^{-1}$ strictly positive definite (we use $\tau=10^{-3}$ in this article), and $\hat G_{\beta,z}$ is an estimate of ${E}(G_{\beta,z})$, the sub-matrix of the negative Fisher information with expectation taken over $\Pi(\beta,z\mid y)$.  In practice, to use its sample counterpart $\hat G_{\beta,z}$, one can take the Hessian matrix evaluated at the posterior mode of $\beta$ and the minimizer $z$,
or the observed Fisher information matrix by averaging samples $G_{\beta,z}$ over the burn-in period of the Markov chain Monte Carlo. In this article, we use the Hessian at the posterior mode.

\begin{remark}
	The constant-value specification of \eqref{eq:precondition} is suitable for problems where the Hessian changes relatively slowly in the high posterior density region. For problems with $G_{\beta,z}$ rapidly changing, the linearization may provide a poor approximation, and thus $G_{\beta,z}^\T M^{-1}$ may be further from zero, rendering \eqref{eq:precondition}  ineffective.
	Broadly speaking, rapidly changing Hessians pose a challenge for 
	existing Hamiltonian Monte Carlo algorithms.
	One remedy involves location-dependent specifications of $M(\theta)$ becomes necessary---it tends to require intensive computation as an alternative to the leapfrog integrator. One can find discussion of those integrators in \citet{girolami2011riemann,lan2015markov,nishimura2016geometrically}. As an alternative, variational inference can be used to bypass the challenges, at the cost of additional posterior approximation.
\end{remark}

\section{Asymptotic Theory}
We now study the asymptotic properties of the proposed method, establishing normality in the large sample limit in a Bernstein--von Mises sense.
In Bayesian analysis, the Bernstein--von Mises theorem asserts that, under mild regularity conditions, the posterior distribution of the centered and scaled maximum likelihood estimator $\sqrt{n}(\beta - \hat{\beta}_n)$ converges as sample size $n$ increases to a normal distribution centered at $0$, with covariance matrix equal to the inverse Fisher information. In models involving a nuisance parameter $z$, the primary focus often lies in the marginal posterior distribution of the parameter of interest, $\beta$ \citep{berger1999integrated, severini2007integrated}. This marginal posterior is related to the complete posterior by integrating out the nuisance variable $z$: that is, $\Pi(\beta \mid y) \propto \int_{\mathcal{H}} L(y, z;\beta) \pi_0(\beta) \,\textup{d} z $. The domain of integration $z \in \mathcal{H}$ need not be Euclidean, in which case the Bayesian model is a semi-parametric one, so that the Bernstein--von Mises theorem provides a feasible way to characterize the asymptotic posterior when the parameter $z$ lies in any Hilbert space. 

\citet{bickel2012semiparametric} established a Bernstein--von Mises theorem for the marginal posterior under the existence of the least favorable model, provided that the posterior concentrates around this model. Building upon this foundation, we establish a Bernstein--von Mises result for the gradient-bridged posterior. Our approach hinges on the condition that the posterior distribution concentrates around the optimal value of the parameter $z$, as determined by the the first-order optimality condition. This is satisfied due to the shrinkage kernel.

We assume that $\beta\in \Theta \subset \mathbb{R}^d$, and $z\in \mathcal{H}$ lies in a Hilbert space. Let $\beta_0,z_0$ be the fixed, ground-truth values. The prior $\pi_0(\beta)$ is assumed to be continuous at $\beta_0$ with $\pi_0(\beta_0)>0$. We define the log-likelihood as
$l_n(\beta, z) = \log L_n(\beta, z; y)$, where the subscript indicates the sample size.
We use $\|\cdot\|$ to denote the Euclidean--Frobenius norm.  We denote the generative distribution of the data by $P_{\beta,z}$ with parameters $\beta$ and $z$.
We use $H(P,Q)$ to denote the Hellinger distance between two probability measure $P$ and $Q$, and define a metric $d_H$ on the space of $z$ by $d_H(z_1, z_2; \beta) = H(P_{\beta, z_1}, P_{\beta, z_2})$.  We denote the unique minimizer of the gradient-bridge function $h(\beta, z; y)$ by $\hat{z}(\beta):=\hat{z}_\beta$.
We now state the sufficient conditions for establishing the Bernstein--von Mises theorem.

\begin{assumption} \label{assump:1-1}
	There exists a decreasing sequence $\{\rho_n\}$ converging to $0$ such that for every bounded sequence $h_n$,  with $\beta_n = \hat{\beta}_n + h_n / \sqrt{n}$, 
	$$
	\Pi[d_H\{z,\hat{z}(\beta); \beta\}>\rho_n \mid \beta = \beta_n; y] \xrightarrow[n\to\infty]{P_{\beta_0,z_0}} 0\,,
	$$
	where the above denotes convergence in probablity with respect to $P_{\beta_0,z_0}$.
\end{assumption}

\begin{assumption} \label{assump:1-2}  
	There exists a symmetric $H_n(z)$ such that for $\beta_n = \hat{\beta}_n + h_n / \sqrt{n}$,
	$$
	l_n(\beta_n, z) = l_n\{\hat{\beta}_n, \hat{z}(\hat{\beta}_n)\} - \frac{1}{2}  h_n^\T H_n(z) h_n + r_n(h_n, z)  
	$$
	holds for all $z$ satisfying $d_H\{z,\hat{z}(\beta_n); \beta_n\} < \delta$, with remainder term  $r_n(h_n, z)  
	= o_{P_{\beta_0, z_0}}(1)$, and $H_n\{\hat{z}(\hat{\beta}_n)\} \to H_0$.
\end{assumption}

\begin{assumption} \label{assump:1-3}
	There exists a constant $C$ such that for $\beta_n = \hat{\beta}_n + h_n / \sqrt{n}$,
	$$
	\sup_{z: d_H\{z, \hat{z}(\beta_n); \beta_n\} < \rho_n } \int L_n(\beta_n, z;y)\, \textup{d} P_{\beta_0,z_0}(y) \leq C.
	$$
\end{assumption}

We refer to the set of models when $z=\hat{z}(\beta)$ as the constrained favorable submodels.
Assumption \ref{assump:1-1} posits that the conditional posterior concentrates around the constrained favorable submodels. One way for this assumption to be satisfied is by letting $\lambda=\lambda_n \to \infty$. Assumption \ref{assump:1-2} is the stochastic local asymptotic normality of the likelihood around the constrained favorable submodels, which is standard for proving Bernstein--von Mises results. In parametric settings when $z$ belongs to a Euclidean space, the condition can often be proven using Taylor expansion, provided conditions on the derivatives of the likelihood function. Assumption \ref{assump:1-3} is the domination condition of the likelihood around the constrained favorable submodels.

Let the integrated likelihood $S_n(\beta) = \int_{\mathcal{H}} L_n(\beta, z; y)\, \textup{d}\Pi_0(z) $, and $s_n(\beta) = \log S_n(\beta)$. We have the following lemma for the locally asymptotically normal property of the integrated likelihood.

\begin{lemma}\label{lemma:LAN}
	Under Assumption \ref{assump:1-1}--\ref{assump:1-3}, for every bounded sequence $h_n$ and $\beta_n=\hat{\beta}_n+h_n/\sqrt{n}$,
	\bel
	s_{n}(\beta_n) = s_{n}(\hat{\beta}_n)  -\frac{1}{2}  h_n^\T H_0 h_n
	+ o_{P_{\beta_0,z_0}}(1). \label{eq:lan}
	\eel
\end{lemma}

With the locally asymptotically normal condition for $s_n(\beta)$, we make the Bernstein--von Mises result statement.
\begin{theorem} \label{thm:BvM_semi}
	Assume \eqref{eq:lan} holds with positive definite $H_0$. Suppose that the maximum likelihood estimator $\hat \beta_n$ exists and converges to $\beta_0$ when $n\to\infty$; there exists $\delta>0$ such that for any sequence of positive numbers $M_n\to \infty$,
	\be
	\Pr_{\beta_0,z_0} \biggl[ \inf_{z\in \mathcal{H}} \inf_{ \sqrt{n} \|\beta - \hat\beta_n \| \geq M_n} \bigl\{ l_n( \hat \beta_n)-l_n(\beta) \bigr\} \ge \frac{\delta M_n^2}{n}  \biggr] \xrightarrow[n\to\infty]{} 1. 
	\ee
	Then letting $\pi_n$ be the density of $\beta$ when $\beta \sim \Pi_n(\beta \mid y)$, we have
	\(
	\int_{B_{\epsilon}(\beta_{0})} \pi_{n}(\beta)\, \textup{d} \beta \xrightarrow[n\to\infty]{P_{\beta_0,z_0}}  1 \text { for all } \epsilon>0.
	\)
	Moreover, letting $q_{n}$ denote the density of $\sqrt{n}(\beta-\hat\beta_{n})$, we have $\text{d}_{\text{TV}}\bigl\{q_n,\mathcal{N}\bigl(0,H_0^{-1}\bigr)\bigr\} \to 0$ 
	in probability with respect to $P_{\beta_0,z_0}$ as $n\to\infty$.
\end{theorem}

\section{Simulation Study}\label{subsec:flow}

\begin{figure}[ht]
	\centering
	\includegraphics[width=.45\textwidth]{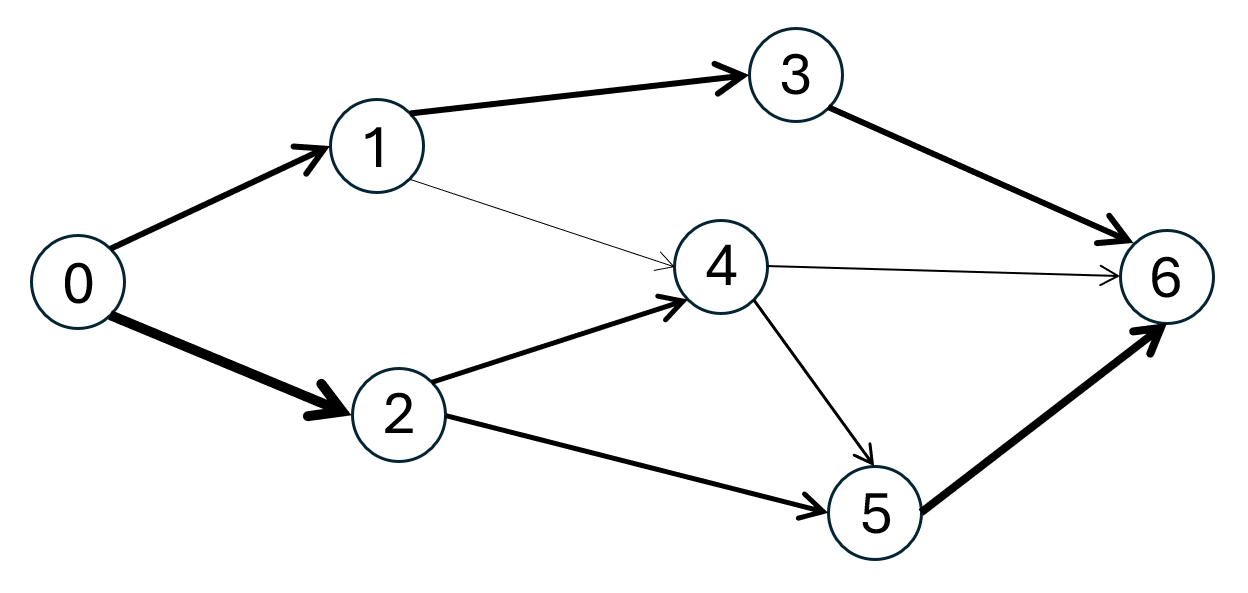}
	\caption{The directed network we use for the experiment on the maximum flow problem. The width of the edges is proportional to the magnitude of the optimal flow.}
	\label{fig:flow-net}
\end{figure}

We empirically illustrate the gradient-bridged posterior with application to the flow network described in Example \ref{example:flow}. The directed network used in the synthetic experiments is shown in Fig. \ref{fig:flow-net}. We use package \texttt{networkx} to generate a flow network, which consists of $7$ nodes and $10$ directed edges. Node $0$ is the source and node $6$ is the sink, and assign ground-truth edge capacities $\beta^0_{ij}$  depicted in Figure \ref{fig:flow-net}. Given this network, we begin by computing the optimal flow values $z^0$ under $\beta^0$ by solving the maximum flow optimization problem. Next, we simulate observed flow data $y^s_{ij}$ for $s=1,\dots,1000$, where each observation $s$ represents a noisy measurement of the optimal flow. We simulate the designed capacity $c_{ij}$ from $\text{No}(\beta_{ij}^0, 0.5^2)$. 

\begin{figure}[ht]
	
	\vspace{0.5cm}
	
	\centering
	\begin{overpic}[width=.43\textwidth]{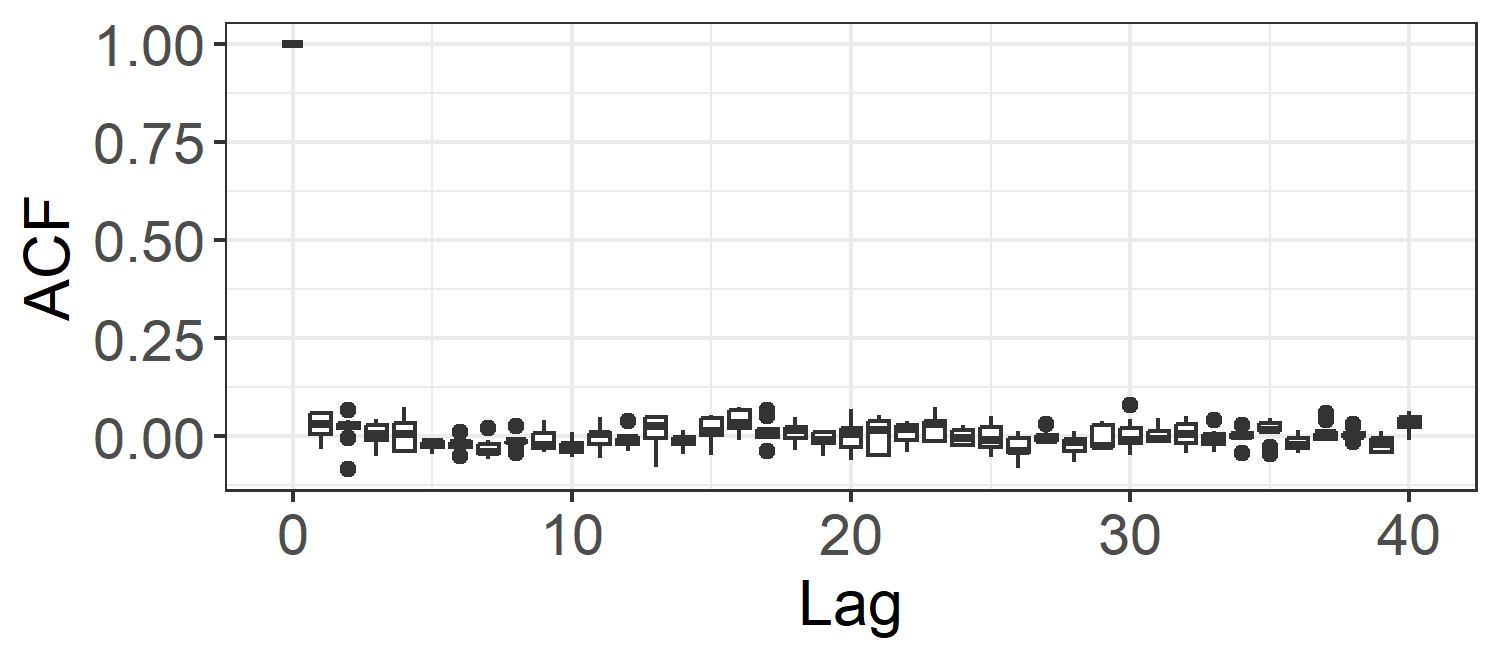}
		\put(15,45){\small (a) $z_{ij}$'s, using our suggested $M^{-1}$ \eqref{eq:precondition}}
	\end{overpic}
	\begin{overpic}[width=.43\textwidth]{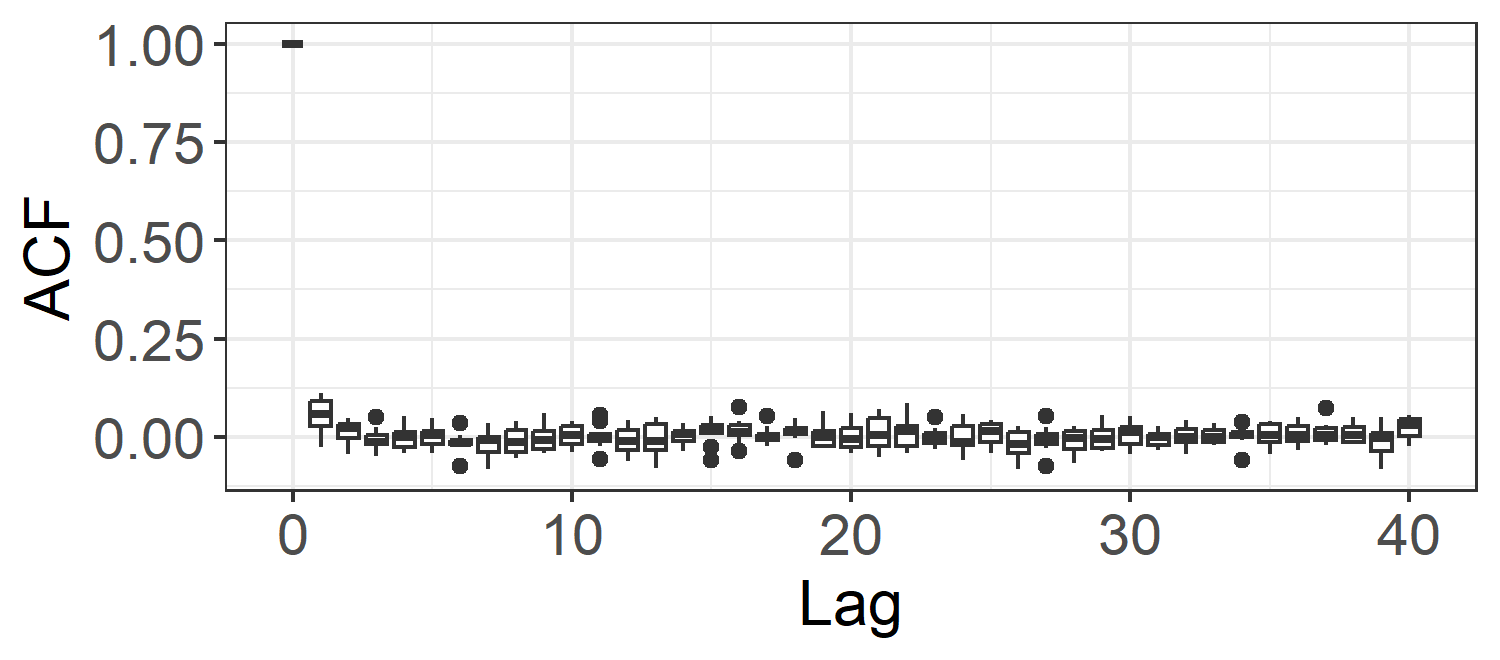}
		\put(15,45){\small (b) $\beta_{ij}$'s, using our suggested $M^{-1}$ \eqref{eq:precondition}}
	\end{overpic}
	
	\vspace{0.5cm}
	
	\centering
	\begin{overpic}[width=.43\textwidth]{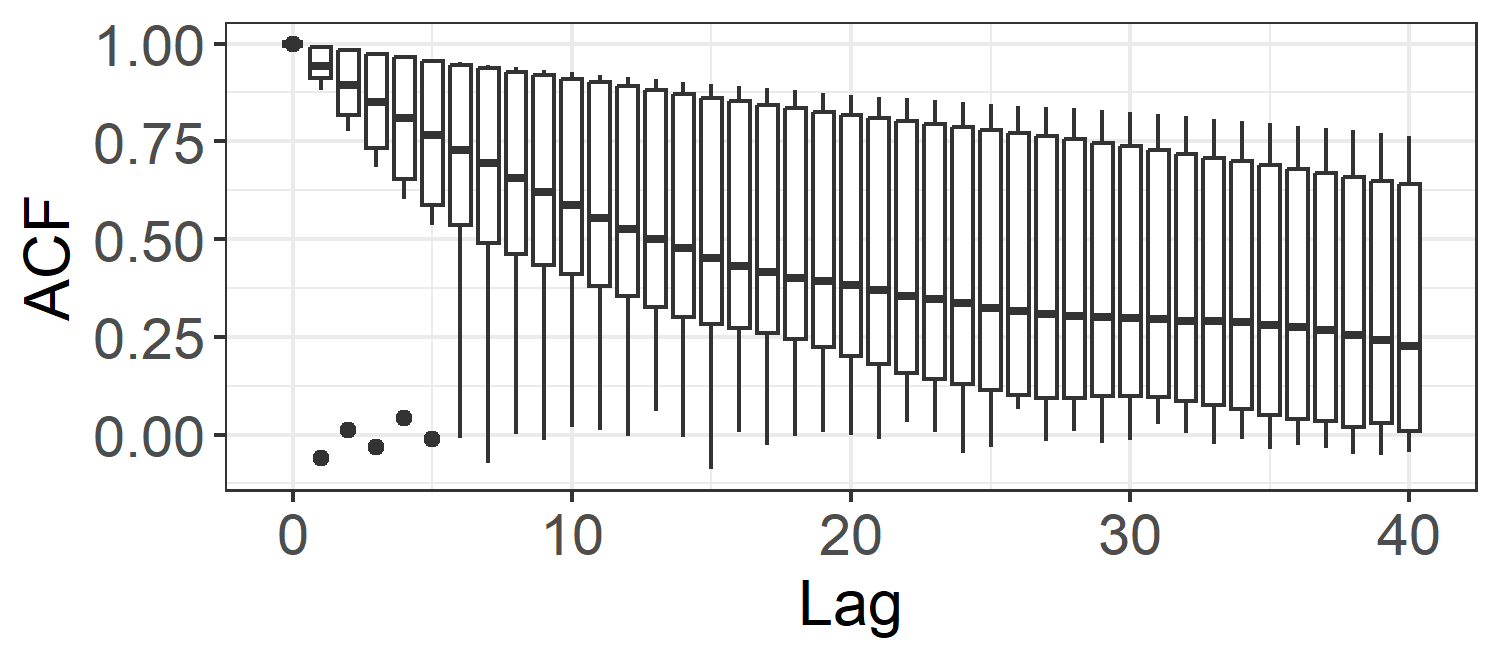}
		\put(15,45){\small (c) $z_{ij}$'s, using default choice of $M^{-1}$}
	\end{overpic}
	\begin{overpic}[width=.43\textwidth]{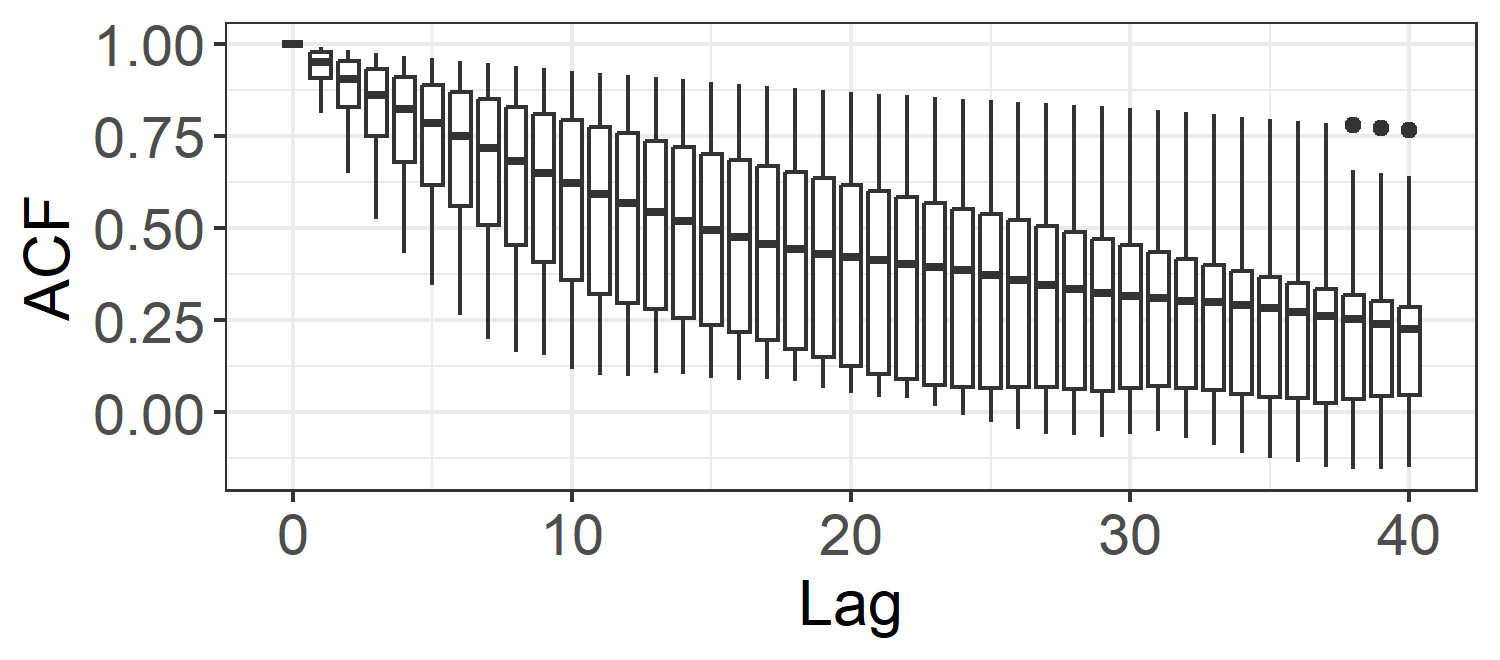}
		\put(15,45){\small (d) $\beta_{ij}$'s, using default choice of $M^{-1}$}
	\end{overpic}
	
	\caption{Autocorrelation of the Markov chain from No-U-Turn Samplers using different choices of the inverse mass matrix. Panels (a) and (b) use our suggested $M^{-1}$ following  \eqref{eq:precondition}. Panels (c) and (d) use the default choice of $M^{-1}$.}
	\label{fig:flow-net-acf}
\end{figure}

We assign independent half-normal $\text{No}_+(0, 10^2)$ priors on the $\beta_{ij}$'s and $z_{ij}$'s, assign $\text{Ga}^{-1}(2,5)$ prior on the $\sigma_y^2$ and assign $\text{Ga}^{-1}(5,2)$ prior on $\sigma_c^2$. Using our suggested choice of  $M^{-1}$ matrix \eqref{eq:precondition}, we use the No-U-Turn sampler \citep{hoffman2014no} to sample from the posterior distribution under \eqref{eq:flowlikelihood} for $12,000$ iterations. The first $2000$ iterations are treated as burn-in, and we apply a thining factor of $10$. Figure \ref{fig:flow-net-acf}(a)--(b) shows the autocorrelation for all coordinates of $z$ and $\beta$; its rapid decay indicates good mixing. In contrast, we also consider the  No-U-Turn Sampler under the default choice of adaptive $M^{-1}$. It is visually clear from panels \ref{fig:flow-net-acf}(c)--(d) that our choice of $M^{-1}$ significantly improves performance in terms of mixing.

\begin{figure}[ht]
	\centering
	\includegraphics[width=.19\textwidth]{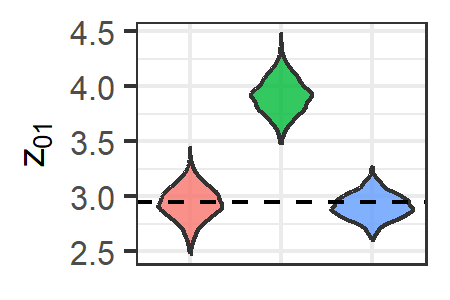}
	\includegraphics[width=.19\textwidth]{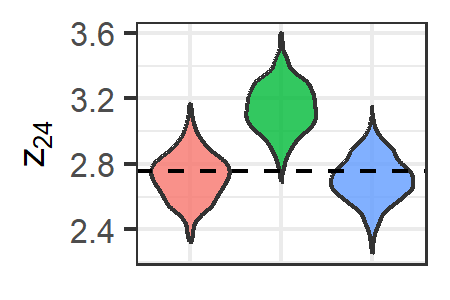}
	\includegraphics[width=.19\textwidth]{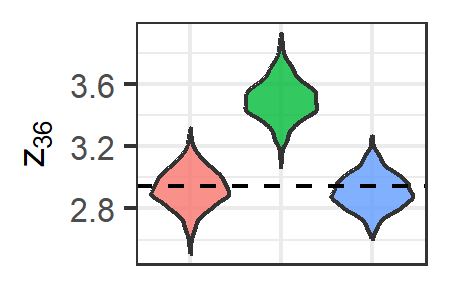}
	\includegraphics[width=.19\textwidth]{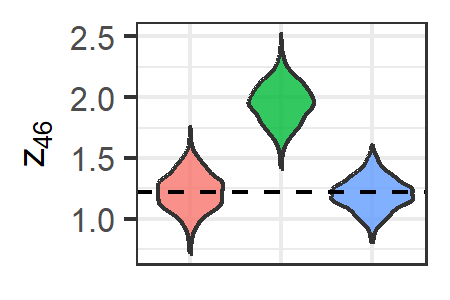}
	\includegraphics[width=.19\textwidth]{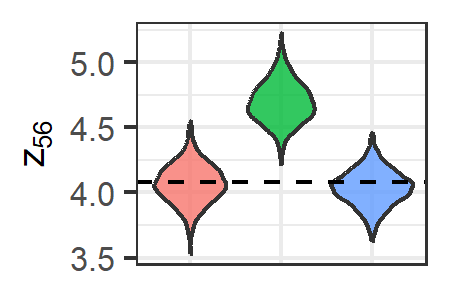}
	\includegraphics[width=.19\textwidth]{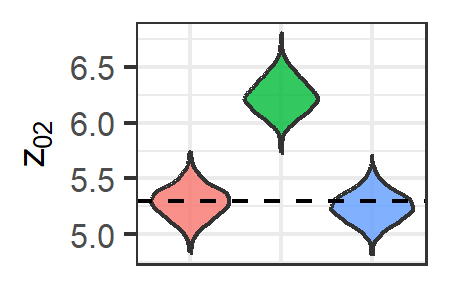}
	\includegraphics[width=.19\textwidth]{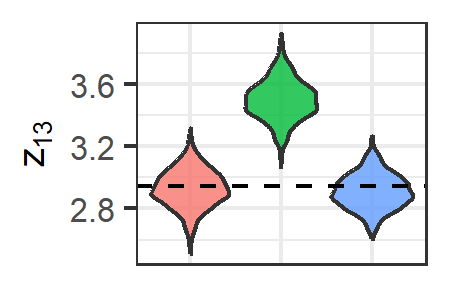}
	\includegraphics[width=.19\textwidth]{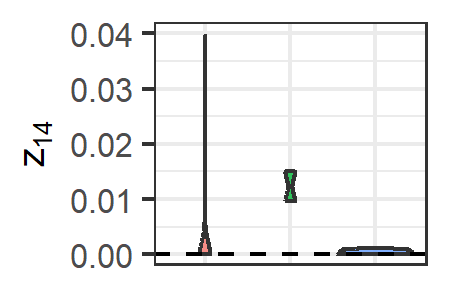}
	\includegraphics[width=.19\textwidth]{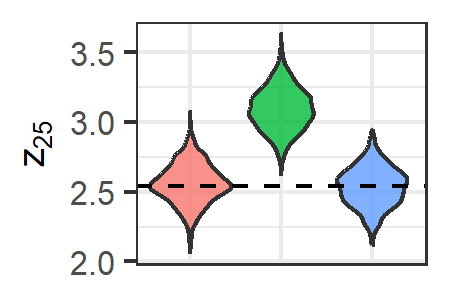}
	\includegraphics[width=.19\textwidth]{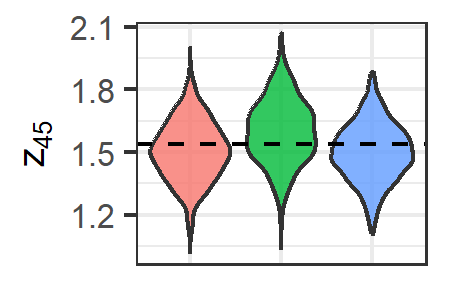}
	\caption{Flow value posteriors $z_{ij}$ for each network edge from the Gibbs posterior with $\lambda=0$ (red), the Gibbs posterior with shrinkage kernel $\exp\{-\lambda h(\beta, z; y)\}$ (green), and the gradient-bridged posterior (blue); horizontal dashed lines depict ground-truth values $z^0_{ij}$.}
	\label{fig:flow-net-z-hist}
\end{figure}

\begin{figure}[ht]
	\centering
	\includegraphics[width=.19\textwidth]{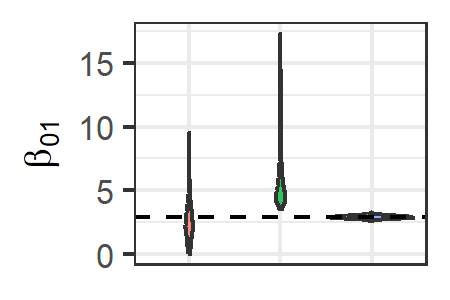}
	\includegraphics[width=.19\textwidth]{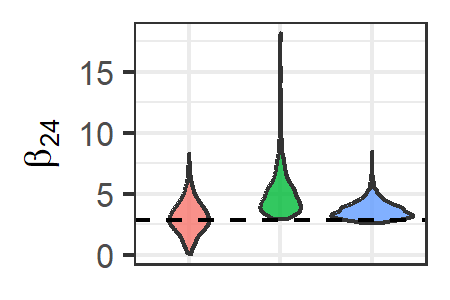}
	\includegraphics[width=.19\textwidth]{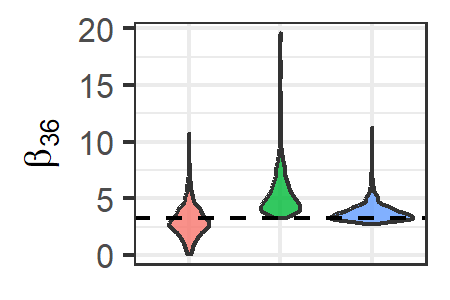}
	\includegraphics[width=.19\textwidth]{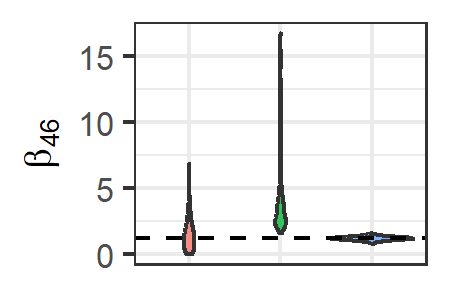}
	\includegraphics[width=.19\textwidth]{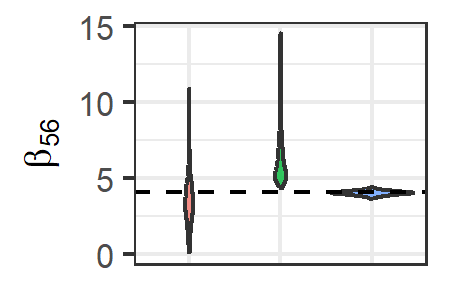}
	\includegraphics[width=.19\textwidth]{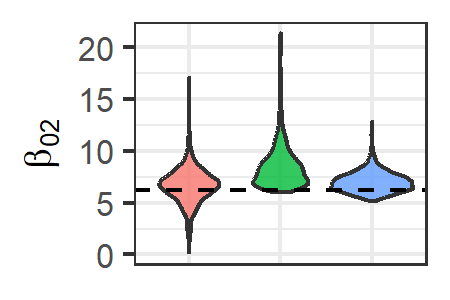}
	\includegraphics[width=.19\textwidth]{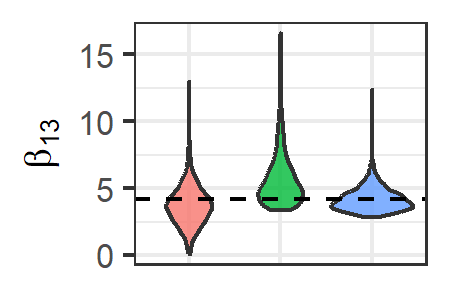}
	\includegraphics[width=.19\textwidth]{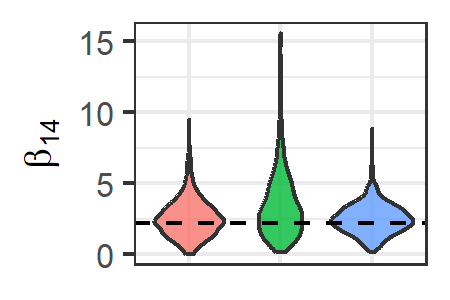}
	\includegraphics[width=.19\textwidth]{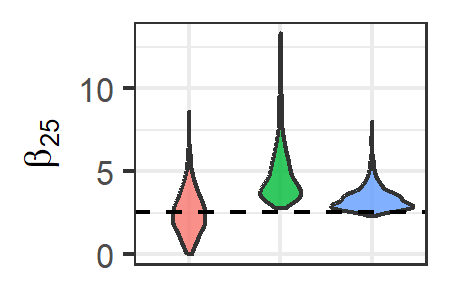}
	\includegraphics[width=.19\textwidth]{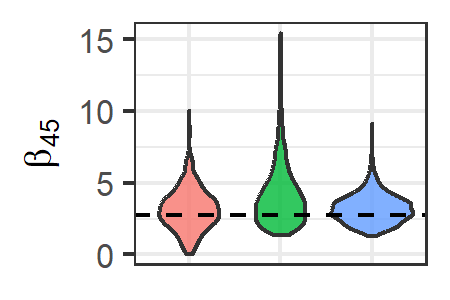}
	\caption{Capacity parameter posteriors $\beta_{ij}$ for each network edge from the Gibbs posterior with $\lambda=0$ (red), the Gibbs posterior with shrinkage kernel $\exp\{-\lambda h(\beta, z; y)\}$ (green), and the gradient-bridged posterior (blue); horizontal dashed lines depict true values $\beta^0_{ij}$.}
	\label{fig:flow-net-beta-hist}
\end{figure}

Figures~\ref{fig:flow-net-z-hist} and \ref{fig:flow-net-beta-hist} present the posterior distributions of the flow variables $z$ and the capacity parameters $\beta$. An immediate takeaway is that empirically, the gradient-bridged posterior is able to recover the ground truth, assigning high posterior mass concentrated around the ground truth $z^0$ and $\beta^0$ values. To highlight the benefits of our proposed method, we compare performance to inference using existing Gibbs posterior approaches. A canonical Gibbs posterior, corresponding to the loss \eqref{eq:flowlikelihood} taking $\lambda=0$, also results in a posterior that assigns mass covering the ground truth values. While estimates of $z$ are quite similar to those under our approach, estimation of $\beta_{ij}$ exhibits strikingly higher variability (Figure \ref{fig:flow-net-beta-hist}). To understand this phenomenon, recall that there are $1000$ noisy flow measurements $y^s$, but only one observed measurement of the capacity $c$. The standard Gibbs posterior approach makes good use of $y$ to estimate $z$, but fails to borrow information in the relationship between $z$ and $\beta$. Instead, our shrinkage kernel exploits the relationship between $\beta$ and $z$ via the implicit function, making the posterior of $\beta$ more accurate.   

To be complete, we additionally consider a Gibbs posterior approach that attempts to account for the relationship between $\beta$ and $z$ via the loss-based kernel
$\exp\{-\lambda h(\beta, z; y)\}$; that is, performing joint shrinkage in lieu of the proposed methodology allowing for \textit{conditional} shrinkage via partial minimization. Empirically, we see that such joint regularization is a poor substitute for accommodating the desired sub-problem. Performance here is arguably even worse, with posteriors now exhibiting bias in both $z$ as well as $\beta$. We see that penalizing $h(\beta,z ; y)$ can lead to poor estimation and inflated posterior variance in both Figures \ref{fig:flow-net-z-hist} and \ref{fig:flow-net-beta-hist}, while the gradient-bridged posterior is free from this issue due to partial minimization over $z$.

Additional simulation results related to the flow network modeling problem are detailed in the Supplement. There we also include another synthetic experiment on the latent quadratic model \citep{zeng2024bridged}, a computationally efficient alternative to the latent Gaussian model, again finding that the gradient-bridged posterior allows for excelleng mixing performance and accurate inference in a relatively high-dimensional setting with $z\in\mathbb{R}^{1000}$.

\section{Data Application: Data Integration for Single-cell Data}

We now consider a case study on data integration task. For data that are reproduced across studies or otherwise collected in batches, study-specific sources of variation may arise artifactually, and should be accounted for to better estimate the desired shared signal. Many Bayesian studies of integrating data across studies take a natural factor analytic approach 
\citep{de2021bayesian,roy2021perturbed,avalos2022heterogeneous,chandra2024inferring}. While some of these studies give special attention to various possible identifiability issues, due to the rotationally invariant nature of the problem formulation, they all require alignment of the estimated loadings via post-processing. These studies have proposed post-hoc alignment via orthogonal procrustes problems as well as Varimax rotation \citep{assmann2016bayesian,rohe2023vintage}.

The gradient-bridged posterior now enables a straightforward way to address the alignment problem directly. We can include an orthogonal procrustes sub-problem in defining our model likelihood 
\citep{gower1975generalized,ma2024principled}. Here, we illustrate how just using a distance-based objective, the methodology allows us to generate aligned samples toward a unified representation that mitigates batch effects while preserving information from cell types. We remark that this machinery can be combined within model-based approaches such as factor models to ameliorate rotational ambiguity.

We consider human pancreatic data \citep{stoeckius2017simultaneous} consisting of transcriptomic profiles from multiple types of human pancreatic cells. These profiles are obtained through $B=5$ batches of sequencing technologies, each with distinct technical biases and variations in sequencing depth. For each batch, the data were processed and transformed into a matrix $X_b\in \bbR^{d\times n}\ (b=1,\dots,B)$, where $d$ is the number of features, $n$ is the sample size; hence $X_{b,(.,i)}$ is the $i$-th observation in the $b$-th batch. The data pre-processing procedures are provided in the Supplementary Materials. To align these batches, we introduce a shared centering parameter $u\in \mathbb{R}^{d\times n}$, a scaling parameter $s_b>0$ for adjusting the magnitude of the data and an approximate rotation-reflection matrix $R_b\in \bbR^{d\times d} \ (i=1,\dots,B)$ for each batch. Following our result described in Example \ref{example:procrustes}, our point of departure is the primal problem
\[\label{eq:proc_data_app}
\min_{R_b} \|R_b X_b- s_b u\|^2_F, \quad \text{subject to } R_b^{\T} R_b=I_d,
\]
over batches $b=1,\ldots, B$. Then, following Equation \eqref{eq:procrutes_shrinkage} we construct the shrinkage kernel  
\(
\exp\{-\lambda \sum_{b=1}^B \|W_b W_b^\T (s_b^2 X_b u^\T u X_b^\T)W_bW_b^\T-I \|^2_F \},
\)
with $R_b= s_b u X_b^\T(W_bW_b^\T)$. This positions us to quantify measure of fit under the remaining contribution to the likelihood \(
g(y;\beta,z)=\prod_{b=1}^n(\sigma^2)^{-(dn)/2} \exp\{-\|R_bX_b- s_b u\|^2_F/(2\sigma^2)\}.\)
At a large $\lambda$, the gradient-bridged posterior drives the gradient of the dual objective nearly to zero, thereby approximately satisfying the orthogonality constraint for $R_b$ while aligning it closely with its conditional optimum. In contrast, if $\lambda=0$ and $R^{\rm T}_b R_b=I_d$ exactly, we would have a Gibbs posterior based on the generalized Procrustes loss. The Gibbs posterior strictly enforces the orthogonality constraint, yet it lacks a strong concentration of $R_b$ near its posterior mode, unless $\sigma^2$ is forced to be near zero. The comparison in the following shows an empirical performance difference.

We choose standard half-normal priors for $s_b$,  $\text{Ga}^{-1}(2, 1)$ prior for $\sigma^2$ and a standard normal prior for $u$. We run the No-U-Turn Sampler for $100,000$ iterations, and discard the first $10,000$ iterations as burn-ins, and we thin the chain at $20$. 
For each sample of $\{u, (R_{b})_{b=1}^B\}$, we can produce a sample of the unified representation (aligned data) $(R_b X_b/s_b)_{b=1}^B$.

To assess batch effect mitigation, we use the batch-associated Davies--Bouldin index \citep{davies1979cluster} which quantifies batch separation relative to within-batch compactness. Lower values indicate better batch mixing, which in our case indicates reduced batch effects and thus more successful data integration. Because one index is associated with each aligned sample, the distribution of Davies-Bouldin indices further allows us to evaluate the variability of integration outcomes. We compare results using the gradient-bridged posterior to the same index evaluated on (i) the raw data, (ii) the aligned data via solving the generalized Procrustes problem under loss function \eqref{eq:proc_data_app} under $u=X_1$, and (iii) the samples from a Gibbs posterior as described above taking $\lambda=0$  \citep[using a Gibbs sampler with the help of the \texttt{rstiefel} package,][]{hoff2021rstiefel}. To facilitate comparison to data under (i) and (ii), we derive a point estimate from samples under the gradient-bridged posterior and the Gibbs posterior under a decision-theoretic framework. Specifically, we select the empirical maximizer of the normalized mutual information computed with respect to the cell-type labels. Normalized mutual information is chosen because its values effectively reflect how well cell-type signals are preserved.

\begin{figure}[ht]
	\centering
	\includegraphics[width=.4\textwidth]{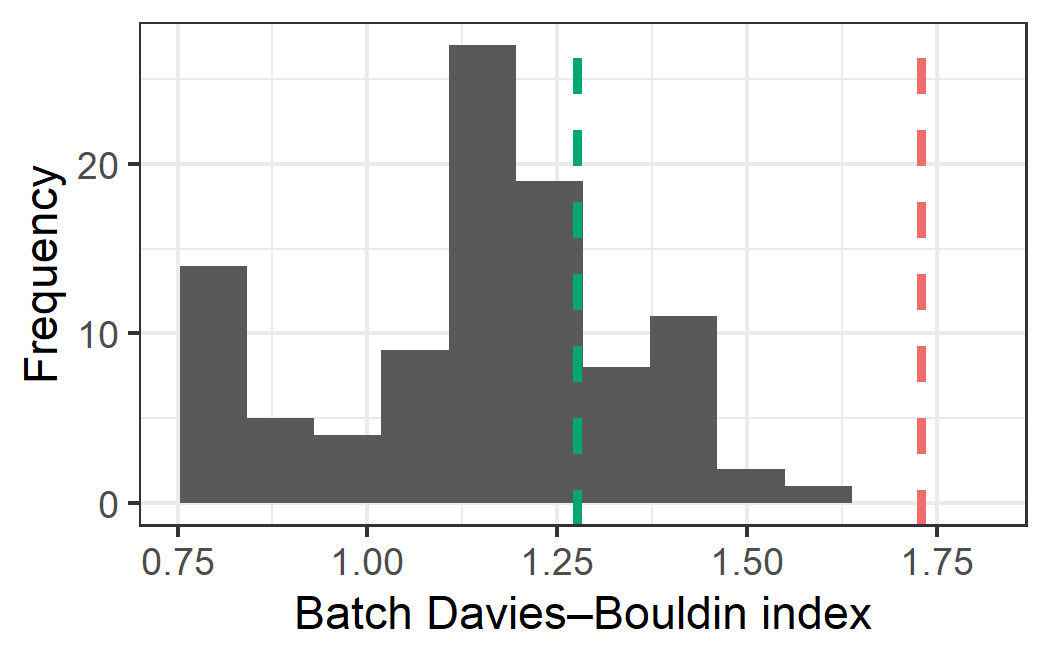}
	\quad
	\includegraphics[width=.4\textwidth]{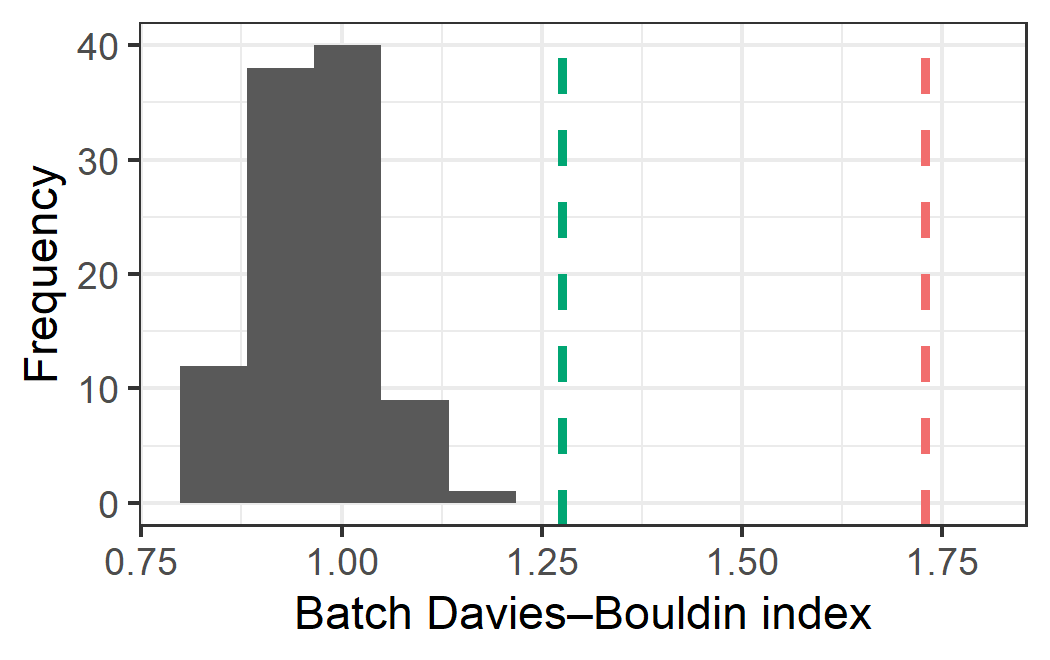}
	\caption{Histograms of posterior samples of batch-associated Davies--Bouldin index, from Gibbs posterior  (left) and gradient-bridged posterior (right). The dashes indicate the corresponding values from the raw data (red) and the generalized Procrustes analysis (green).}
	\label{fig:histograms}
\end{figure}

\begin{figure}[ht]
	\centering
	\includegraphics[width=.24\textwidth]{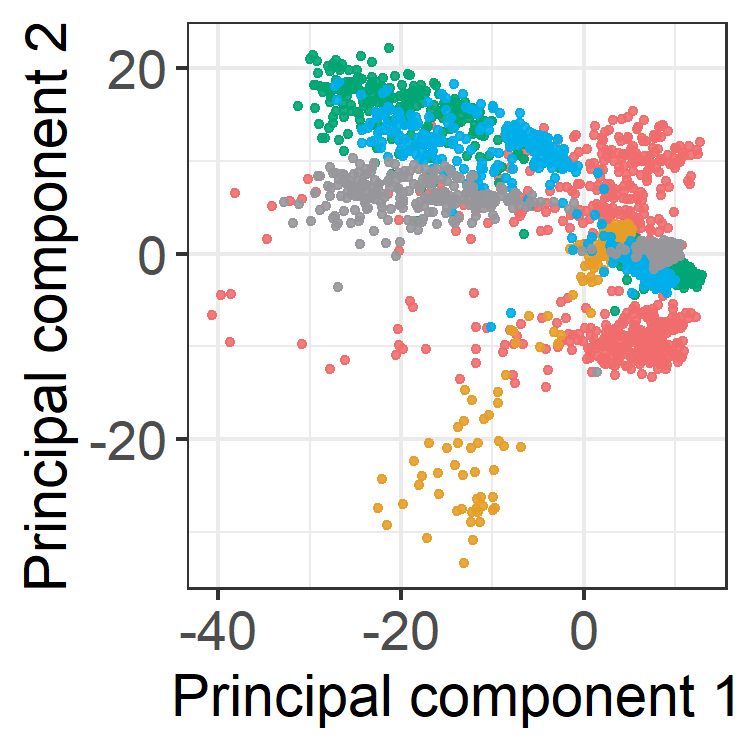}
	\includegraphics[width=.24\textwidth]{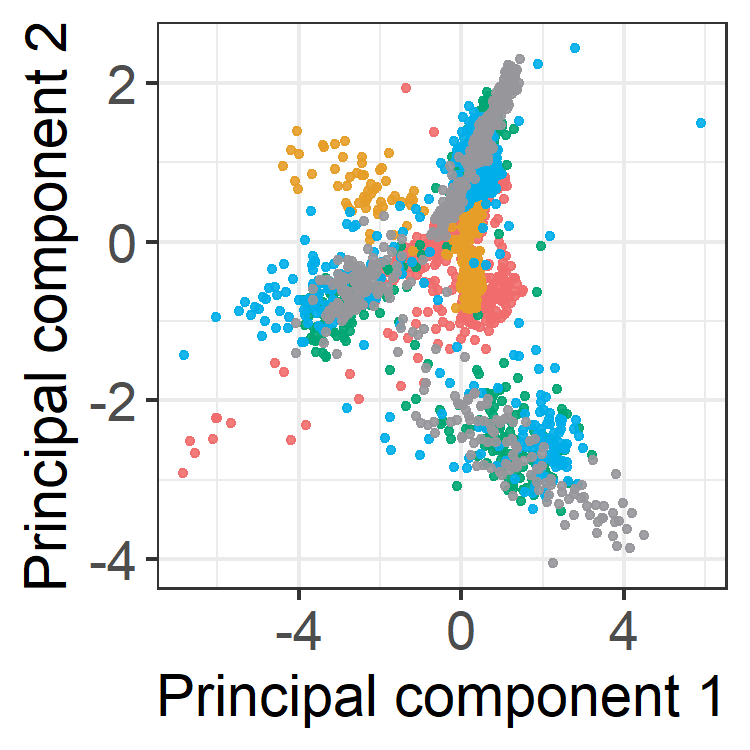}
	\includegraphics[width=.24\textwidth]{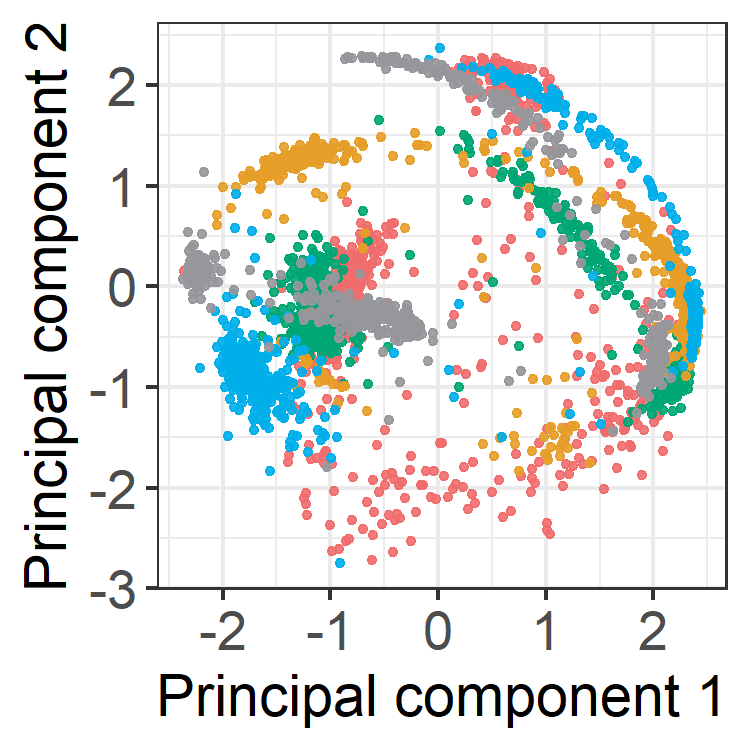}
	\includegraphics[width=.24\textwidth]{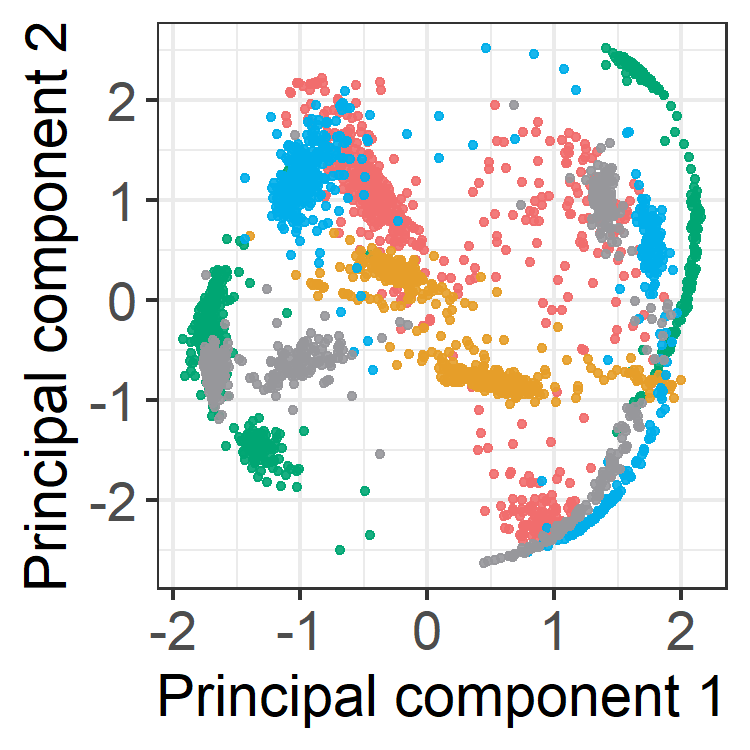}\\
	
	\vspace{0.2cm}
	
	\includegraphics[width=.3\textwidth]{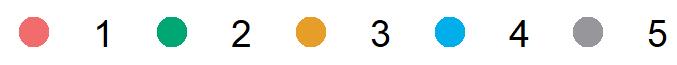}
	\caption{Integrated data represented in two-dimension using principle component analysis colored by batch. Left to right: raw data, generalized Procrustes analysis, Gibbs posterior, gradient-bridged posterior.}
	\label{fig:batch}
\end{figure}

Figure~\ref{fig:histograms} displays the posterior distributions of the batch-associated Davies--Bouldin index. The gradient-bridged posterior shows a substantial improvement in the batch-associated Davies--Bouldin index compared to raw data and other two methods. To visualize the aligned data, we embed the point estimates of the unified representations into the space of the first two principal components. In Figure \ref{fig:batch}, we can see clear batch effects from the raw data and those processed using generalized Procrustes analysis, as the batches, labeled by color, can still be spotted as fairly distinct clusters. The Gibbs posterior improves the batch mixing, but arguably still exhibits noticeable batch separation. The gradient-bridged posterior achieves the best batch mixing. In more detail, these findings are supported quantitatively as presented in Table \ref{tab:metrics_comparison}.

\begin{table}[ht]
	\centering
	\caption{Comparison of point estimates of unified representation in terms of batch-associated Davies--Bouldin index}
	\begin{tabular}{lc}
		\toprule
		Data representation             & Davies--Bouldin Index \\
		\midrule
		Raw data                            & $1.729$   \\ 
		Generalized Procrustes analysis     & $1.276$   \\ 
		Gibbs posterior                     & $1.165$   \\ 
		Gradient-bridged posterior          & $0.965$   \\ 
		\bottomrule
	\end{tabular}
	\label{tab:metrics_comparison}
\end{table}

\section{Discussion}
In this article, we introduce an intuitive approach to account for sub-problems within a Bayesian statistical model. The idea of using a continuous shrinkage kernel on the partial gradient of the sub-problem loss is designed to yield straightforward inference and computational routines. Rather than requiring bespoke samplers to handle constraints or optimization sub-problems, the methodology is designed so that standard gradient-based posterior sampling can be readily applied. The methodology is supported via asymptotic theory, and its performance in finite samples is supported empirically through simulation and case studies. 

While this method is equipped to handle a broad variety of problems, it is currently limited to differentiable sub-problem losses $h$. A natural avenue for future work, then, would extend these ideas to non-differentiable losses such as hinge loss. Proximal mappings provide one route for handling such cases \citep{polson2015proximal}, and recent ideas in the Bayesian literature for handling constraints and priors using these optimization-centric ideas \citep{heng2023bayesian,zhou2024proximal} are likely to be fruitful in similarly extending the gradient-bridged framework. Additionally, for non-convex problems such as phase retrieval \citep{17-EJS1378SI},  it may be advantageous to solve a convex relaxation of the problem while incorporating additional priors to keep $z$ near the approximate solution.

\bibliographystyle{plainnat}
\bibliography{ref}

\begin{thebibliography}{49}
\providecommand{\natexlab}[1]{#1}
\providecommand{\url}[1]{\texttt{#1}}
\expandafter\ifx\csname urlstyle\endcsname\relax
  \providecommand{\doi}[1]{doi: #1}\else
  \providecommand{\doi}{doi: \begingroup \urlstyle{rm}\Url}\fi

\bibitem[A{\ss}mann et~al.(2016)A{\ss}mann, Boysen-Hogrefe, and Pape]{assmann2016bayesian}
Christian A{\ss}mann, Jens Boysen-Hogrefe, and Markus Pape.
\newblock Bayesian analysis of static and dynamic factor models: An ex-post approach towards the rotation problem.
\newblock \emph{Journal of Econometrics}, 192\penalty0 (1):\penalty0 190--206, 2016.

\bibitem[Astfalck et~al.(2024)Astfalck, Sen, Patra, Cripps, and Dunson]{sen2018constrained}
Lachlan Astfalck, Deborshee Sen, Sayan Patra, Edward Cripps, and David Dunson.
\newblock Posterior projection for inference in constrained spaces.
\newblock \emph{arXiv:1812.05741v5}, 2024.

\bibitem[Avalos-Pacheco et~al.(2022)Avalos-Pacheco, Rossell, and Savage]{avalos2022heterogeneous}
Alejandra Avalos-Pacheco, David Rossell, and Richard~S Savage.
\newblock Heterogeneous large datasets integration using {Bayesian} factor regression.
\newblock \emph{Bayesian Analysis}, 17\penalty0 (1):\penalty0 33--66, 2022.

\bibitem[Bahmani and Romberg(2017)]{17-EJS1378SI}
Sohail Bahmani and Justin Romberg.
\newblock A flexible convex relaxation for phase retrieval.
\newblock \emph{Electronic Journal of Statistics}, 11\penalty0 (2):\penalty0 5254--5281, 2017.

\bibitem[Bazaraa et~al.(2011)Bazaraa, Jarvis, and Sherali]{bazaraa2011linear}
Mokhtar~S Bazaraa, John~J Jarvis, and Hanif~D Sherali.
\newblock \emph{Linear programming and network flows}.
\newblock John Wiley \& Sons, 2011.

\bibitem[Berger et~al.(1999)Berger, Liseo, and Wolpert]{berger1999integrated}
James~O Berger, Brunero Liseo, and Robert~L Wolpert.
\newblock Integrated likelihood methods for eliminating nuisance parameters.
\newblock \emph{Statistical Science}, 14\penalty0 (1):\penalty0 23--25, 1999.

\bibitem[Bhattacharya and Martin(2022)]{bhattacharya2022gibbs}
Indrabati Bhattacharya and Ryan Martin.
\newblock Gibbs posterior inference on multivariate quantiles.
\newblock \emph{Journal of Statistical Planning and Inference}, 218:\penalty0 106--121, 2022.

\bibitem[Bickel and Kleijn(2012)]{bickel2012semiparametric}
P.~J. Bickel and B.~J.~K. Kleijn.
\newblock The semiparametric {Bernstein--von} {Mises} theorem.
\newblock \emph{The Annals of Statistics}, 40\penalty0 (1):\penalty0 206--237, 2012.

\bibitem[Bissiri et~al.(2016)Bissiri, Holmes, and Walker]{bissiri2016general}
Pier~Giovanni Bissiri, Chris~C Holmes, and Stephen~G Walker.
\newblock A general framework for updating belief distributions.
\newblock \emph{Journal of the Royal Statistical Society Series B: Statistical Methodology}, 78\penalty0 (5):\penalty0 1103--1130, 2016.

\bibitem[Chakraborty and Ghosal(2022)]{chakraborty2022rates}
Moumita Chakraborty and Subhashis Ghosal.
\newblock Rates and coverage for monotone densities using projection-posterior.
\newblock \emph{Bernoulli}, 28\penalty0 (2):\penalty0 1093--1119, 2022.

\bibitem[Chandra et~al.(2024)Chandra, Dunson, and Xu]{chandra2024inferring}
Noirrit~Kiran Chandra, David~B Dunson, and Jason Xu.
\newblock Inferring covariance structure from multiple data sources via subspace factor analysis.
\newblock \emph{Journal of the American Statistical Association (in press)}, pages 1--15, 2024.

\bibitem[Cheng and Kosorok(2008)]{cheng2008higher}
Guang Cheng and Michael~R Kosorok.
\newblock Higher order semiparametric frequentist inference with the profile sampler.
\newblock \emph{The Annals of Statistics}, 36\penalty0 (4):\penalty0 1786--1818, 2008.

\bibitem[Cheng and Kosorok(2009)]{cheng2009penalized}
Guang Cheng and Michael~R Kosorok.
\newblock The penalized profile sampler.
\newblock \emph{Journal of Multivariate Analysis}, 100\penalty0 (3):\penalty0 345--362, 2009.

\bibitem[Davies and Bouldin(1979)]{davies1979cluster}
David~L Davies and Donald~W Bouldin.
\newblock A cluster separation measure.
\newblock \emph{IEEE Transactions on Pattern Analysis and Machine Intelligence}, PAMI-1\penalty0 (2):\penalty0 224--227, 1979.

\bibitem[De~Vito et~al.(2021)De~Vito, Bellio, Trippa, and Parmigiani]{de2021bayesian}
Roberta De~Vito, Ruggero Bellio, Lorenzo Trippa, and Giovanni Parmigiani.
\newblock Bayesian multistudy factor analysis for high-throughput biological data.
\newblock \emph{The Annals of Applied Statistics}, 15\penalty0 (4):\penalty0 1723--1741, 2021.

\bibitem[Dryden and Mardia(2016)]{dryden2016statistical}
Ian~L Dryden and Kanti~V Mardia.
\newblock \emph{Statistical shape analysis: with applications in R}.
\newblock John Wiley \& Sons, 2016.

\bibitem[Duan et~al.(2020)Duan, Young, Nishimura, and Dunson]{duan2020}
Leo~L Duan, Alexander~L Young, Akihiko Nishimura, and David~B Dunson.
\newblock Bayesian constraint relaxation.
\newblock \emph{Biometrika}, 107\penalty0 (1):\penalty0 191--204, 2020.

\bibitem[Gelfand et~al.(1992)Gelfand, Smith, and Lee]{gelfand1992bayesian}
Alan~E Gelfand, Adrian~FM Smith, and Tai-Ming Lee.
\newblock Bayesian analysis of constrained parameter and truncated data problems using {Gibbs} sampling.
\newblock \emph{Journal of the American Statistical Association}, 87\penalty0 (418):\penalty0 523--532, 1992.

\bibitem[Girolami and Calderhead(2011)]{girolami2011riemann}
Mark Girolami and Ben Calderhead.
\newblock Riemann manifold {Langevin} and {Hamiltonian} {Monte} {Carlo} methods.
\newblock \emph{Journal of the Royal Statistical Society Series B: Statistical Methodology}, 73\penalty0 (2):\penalty0 123--214, 2011.

\bibitem[Goodall(1991)]{goodall1991procrustes}
Colin Goodall.
\newblock Procrustes methods in the statistical analysis of shape.
\newblock \emph{Journal of the Royal Statistical Society Series B: Statistical Methodology}, 53\penalty0 (2):\penalty0 285--321, 1991.

\bibitem[Gower(1975)]{gower1975generalized}
John~C Gower.
\newblock Generalized {Procrustes} analysis.
\newblock \emph{Psychometrika}, 40\penalty0 (1):\penalty0 33--51, 1975.

\bibitem[Heng et~al.(2023)Heng, Zhou, and Chi]{heng2023bayesian}
Qiang Heng, Hua Zhou, and Eric~C Chi.
\newblock Bayesian trend filtering via proximal {Markov} chain {Monte Carlo}.
\newblock \emph{Journal of Computational and Graphical Statistics}, 32\penalty0 (3):\penalty0 938--949, 2023.

\bibitem[Hoff and Franks(2021)]{hoff2021rstiefel}
Peter Hoff and Alexander Franks.
\newblock \emph{rstiefel: Random orthonormal matrix generation and optimization on the Stiefel manifold}, 2021.
\newblock URL \url{https://CRAN.R-project.org/package=rstiefel}.
\newblock R package version 1.0.1.

\bibitem[Hoffman et~al.(2014)Hoffman, Gelman, et~al.]{hoffman2014no}
Matthew~D Hoffman, Andrew Gelman, et~al.
\newblock The {No-U-Turn Sampler}: Adaptively setting path lengths in {Hamiltonian Monte Carlo}.
\newblock \emph{Journal of Machine Learning Research}, 15\penalty0 (1):\penalty0 1593--1623, 2014.

\bibitem[Jiang and Tanner(2008)]{jiang2008gibbs}
Wenxin Jiang and Martin~A Tanner.
\newblock Gibbs posterior for variable selection in high-dimensional classification and data mining.
\newblock \emph{The Annals of Statistics}, 36\penalty0 (5):\penalty0 2207--2231, 2008.

\bibitem[Lan et~al.(2015)Lan, Stathopoulos, Shahbaba, and Girolami]{lan2015markov}
Shiwei Lan, Vasileios Stathopoulos, Babak Shahbaba, and Mark Girolami.
\newblock {Markov} chain {Monte Carlo} from {Lagrangian} dynamics.
\newblock \emph{Journal of Computational and Graphical Statistics}, 24\penalty0 (2):\penalty0 357--378, 2015.

\bibitem[Lee et~al.(2005)Lee, Kosorok, and Fine]{lee2005profile}
Bee~Leng Lee, Michael~R Kosorok, and Jason~P Fine.
\newblock The profile sampler.
\newblock \emph{Journal of the American Statistical Association}, 100\penalty0 (471):\penalty0 960--969, 2005.

\bibitem[Lee et~al.(2023)Lee, Lee, and Lee]{lee2023post}
Kwangmin Lee, Kyoungjae Lee, and Jaeyong Lee.
\newblock Post-processed posteriors for banded covariances.
\newblock \emph{Bayesian Analysis}, 18\penalty0 (3):\penalty0 1017--1040, 2023.

\bibitem[Ma et~al.(2024)Ma, Sun, Donoho, and Zou]{ma2024principled}
Rong Ma, Eric~D Sun, David Donoho, and James Zou.
\newblock Principled and interpretable alignability testing and integration of single-cell data.
\newblock \emph{Proceedings of the National Academy of Sciences}, 121\penalty0 (10):\penalty0 e2313719121, 2024.

\bibitem[Maclaren(2018)]{maclaren2018profile}
Oliver~J Maclaren.
\newblock Is profile likelihood a true likelihood? {An} argument in favor.
\newblock \emph{arXiv:1801.04369v4}, 2018.

\bibitem[Martin and Syring(2022)]{martin2022direct}
Ryan Martin and Nicholas Syring.
\newblock Direct {Gibbs} posterior inference on risk minimizers: Construction, concentration, and calibration.
\newblock In \emph{Handbook of Statistics}, volume~47, pages 1--41. Elsevier, 2022.

\bibitem[Miller(2021)]{miller2021asymptotic}
Jeffrey~W Miller.
\newblock Asymptotic normality, concentration, and coverage of generalized posteriors.
\newblock \emph{Journal of Machine Learning Research}, 22\penalty0 (168):\penalty0 1--53, 2021.

\bibitem[Murphy and Van~der Vaart(2000)]{murphy2000profile}
Susan~A Murphy and Aad~W Van~der Vaart.
\newblock On profile likelihood.
\newblock \emph{Journal of the American Statistical Association}, 95\penalty0 (450):\penalty0 449--465, 2000.

\bibitem[Natarajan et~al.(2024)Natarajan, De~Iorio, Heinecke, Mayer, and Glenn]{natarajan2024cohesion}
Abhinav Natarajan, Maria De~Iorio, Andreas Heinecke, Emanuel Mayer, and Simon Glenn.
\newblock Cohesion and repulsion in {Bayesian} distance clustering.
\newblock \emph{Journal of the American Statistical Association}, 119\penalty0 (546):\penalty0 1374--1384, 2024.

\bibitem[Neal(2011)]{neal2011mcmc}
Radford~M Neal.
\newblock {MCMC} using {Hamiltonian} dynamics.
\newblock In \emph{Handbook of Markov Chain Monte Carlo}, pages 113--162. Chapman and Hall/CRC, 2011.

\bibitem[Nishimura and Dunson(2017)]{nishimura2016geometrically}
Akihiko Nishimura and David Dunson.
\newblock Geometrically tempered {Hamiltonian} {Monte} {Carlo}.
\newblock \emph{arXiv:1604.00872v2}, 2017.

\bibitem[Polson and Scott(2016)]{polson2016mixtures}
Nicholas~G Polson and James~G Scott.
\newblock Mixtures, envelopes and hierarchical duality.
\newblock \emph{Journal of the Royal Statistical Society Series B: Statistical Methodology}, 78\penalty0 (4):\penalty0 701--727, 2016.

\bibitem[Polson et~al.(2015)Polson, Scott, and Willard]{polson2015proximal}
Nicholas~G Polson, James~G Scott, and Brandon~T Willard.
\newblock Proximal algorithms in statistics and machine learning.
\newblock \emph{Statistical Science}, 30\penalty0 (4):\penalty0 559--581, 2015.

\bibitem[Presman and Xu(2023)]{presman2023}
Rick Presman and Jason Xu.
\newblock Distance-to-set priors and constrained {Bayesian} inference.
\newblock In \emph{Proceedings of The 26th International Conference on Artificial Intelligence and Statistics}, volume 206 of \emph{Proceedings of Machine Learning Research}, pages 2310--2326. PMLR, 2023.

\bibitem[Rigon et~al.(2023)Rigon, Herring, and Dunson]{rigon2023generalized}
Tommaso Rigon, Amy~H Herring, and David~B Dunson.
\newblock A generalized {Bayes} framework for probabilistic clustering.
\newblock \emph{Biometrika}, 110\penalty0 (3):\penalty0 559--578, 2023.

\bibitem[Rohe and Zeng(2023)]{rohe2023vintage}
Karl Rohe and Muzhe Zeng.
\newblock Vintage factor analysis with {Varimax} performs statistical inference.
\newblock \emph{Journal of the Royal Statistical Society Series B: Statistical Methodology}, 85\penalty0 (4):\penalty0 1037--1060, 2023.

\bibitem[Roy et~al.(2021)Roy, Lavine, Herring, and Dunson]{roy2021perturbed}
Arkaprava Roy, Isaac Lavine, Amy~H Herring, and David~B Dunson.
\newblock Perturbed factor analysis: Accounting for group differences in exposure profiles.
\newblock \emph{The Annals of Applied Statistics}, 15\penalty0 (3):\penalty0 1386--1404, 2021.

\bibitem[Severini(2007)]{severini2007integrated}
Thomas~A Severini.
\newblock Integrated likelihood functions for non-{Bayesian} inference.
\newblock \emph{Biometrika}, 94\penalty0 (3):\penalty0 529--542, 2007.

\bibitem[Stoeckius et~al.(2017)Stoeckius, Hafemeister, Stephenson, Houck-Loomis, Chattopadhyay, Swerdlow, Satija, and Smibert]{stoeckius2017simultaneous}
Marlon Stoeckius, Christoph Hafemeister, William Stephenson, Brian Houck-Loomis, Pratip~K Chattopadhyay, Harold Swerdlow, Rahul Satija, and Peter Smibert.
\newblock Simultaneous epitope and transcriptome measurement in single cells.
\newblock \emph{Nature Methods}, 14\penalty0 (9):\penalty0 865--868, 2017.

\bibitem[Tanner and Wong(1987)]{tanner1987calculation}
Martin~A Tanner and Wing~Hung Wong.
\newblock The calculation of posterior distributions by data augmentation.
\newblock \emph{{Journal of the American Statistical Association}}, 82\penalty0 (398):\penalty0 528--540, 1987.

\bibitem[Van~Dyk and Meng(2001)]{van2001}
David~A Van~Dyk and Xiao-Li Meng.
\newblock The art of data augmentation.
\newblock \emph{Journal of Computational and Graphical Statistics}, 10\penalty0 (1):\penalty0 1--50, 2001.

\bibitem[West(2024)]{west2024perspectives}
Mike West.
\newblock Perspectives on constrained forecasting.
\newblock \emph{Bayesian Analysis}, 19\penalty0 (4):\penalty0 1013--1039, 2024.

\bibitem[Zeng et~al.(2024)Zeng, Dilma, Xu, and Duan]{zeng2024bridged}
Cheng Zeng, Eleni Dilma, Jason Xu, and Leo~L Duan.
\newblock The bridged posterior: optimization, profile likelihood and a new approach to generalized {Bayes}.
\newblock \emph{arXiv:2403.00968}, 2024.

\bibitem[Zhou et~al.(2024)Zhou, Heng, Chi, and Zhou]{zhou2024proximal}
Xinkai Zhou, Qiang Heng, Eric~C Chi, and Hua Zhou.
\newblock Proximal {MCMC} for {B}ayesian inference of constrained and regularized estimation.
\newblock \emph{The American Statistician}, 78\penalty0 (4):\penalty0 379--390, 2024.

\end{thebibliography}

\appendix

\section{Proofs}

\begin{proof}[of Theorem 1]
	Let $A_0:= \nabla^2_{zz} h(\beta_0,z_0; y)$, and consider a mapping:
	$$
	G(\beta,z) =z - A_0^{-1} \nabla_z h(\beta,z;y),
	$$
	for which a fixed point $z$ of $G(\beta,\cdot)$ satisfies $G(\beta,z)=z \Rightarrow \nabla_z h(\beta,z;y)=0$. We can now show $G(\beta, \cdot)$ is a contraction for $z$ inside a neighborhood of $z_0$ defined via the Hessian $\nabla^2_{zz} h(\beta,z; y)$.
	
	Since $\nabla_z G(\beta_0,z_0)=I-A_0^{-1}A_0=0$, by the continuity of $\nabla^2_{zz} h(\beta, z; y)$, we can find $z$ in a neighborhood of $z_0$, $\mathbb{B}(z_0,k; \beta_0):= \{z:\|\nabla_z G(\beta_0,z)\|_{op}=
	\|I-A_0^{-1}\nabla^2_{zz} h(\beta_0,z; y) \|_{op}
	\le k<1\}$, by the mean value theorem, $G(\beta_0,\cdot)$ is a contraction in $\mathbb{B}(z_0,k; \beta_0)$, i.e. $\|G(\beta_0,z_0)-G(\beta_0,z)\| \le k\|z_0-z\|$. It follows that
	\(
	\| z-z_0\| &= \|G(\beta_0,z) -  G(\beta_0,z_0)  + z - G(\beta_0,z)\| \\
	&\le  k \|z-z_0\| + \|z - G(\beta_0,z)\| \\
	&=  k \|z-z_0\| + \|A_0^{-1} \nabla_z h(\beta,z;y)\| \\
	& \le   k \|z-z_0\| + \|A_0^{-1}\|_{op} \epsilon \\
	& =    k \|z-z_0\| +   \lambda^{-1}_{\min}(A_0)\epsilon.
	\)
	Rearranging terms yields the result.
\end{proof}

\begin{proof}[of Lemma 1]
	Let $\eta,\gamma$ be any given positive numbers.
	By the Assumption 1, there exists a large enough $N$ such that $$P_{\beta_0,z_0}\Bigl( \Pi[d_H\{z,\hat{z}(\beta); \beta\} < \rho_n \mid \beta = \beta_n; y] > e^{-\eta} \Bigr) > 1- \gamma$$
	for all $n>N$. Let $D(\beta_n, \rho_n) = \{z: d_H\{z,\hat{z}(\beta_n); \beta_n\} < \rho_n\}$.
	Note that $\Pi(z \mid \beta = \beta_n) = L_n(\beta_n, z;y) / \int_{\mathcal{H}} L_n(\beta_n,z;y)\, \textup{d}\Pi_0(z) = L_n(\beta_n,z;y) / S_n(\beta_n)$, we have $\Pi[d_H\{z,\hat{z}(\beta); \beta\} < \rho_n \mid \beta = \beta_n; y] =  \int_{D(\beta_n, \rho_n)} L_n(\beta_n,z;y)\, \textup{d}\Pi_0(z) / S_n(\beta_n)$. Hence, 
	$$P_{\beta_0,z_0}\biggl( \log \int_{D(\beta_n, \rho_n)} L_n(\beta_n,z;y)\, \textup{d}\Pi_0(z) - \log S_n(\beta_n) > -\eta \biggr) > 1- \gamma\,.$$
	We only need to prove $\log \int_{D(\beta_n, \rho_n)} L_n(\beta_n,z;y)\, \textup{d}\Pi_0(z)$ satisfies the conclusion of Lemma 1.
	
	We define the events $F_n(z, \epsilon) = \{\sup_{h_n} |h_n^\T H_n(z) h_n - h_n^\T H_0 h_n | \leq \epsilon\}$. Then
	\(
	& \int_{D(\beta_n, \rho_n)} L_n(\beta_n,z;y) \, \textup{d}\Pi_0(z) \\
	& = \int_{D(\beta_n, \rho_n)} L_n(\beta_n,z;y) 1_{F_n(z, \epsilon)} \, \textup{d}\Pi_0(z) + \int_{D(\beta_n, \rho_n)} L_n(\beta_n,z;y) 1_{F^c_n(z, \epsilon)} \, \textup{d}\Pi_0(z)\,.
	\)
	Now, the integral of the second term
	\(
	&  \int \int_{D(\beta_n, \rho_n)} L_n(\beta_n,z;y) 1_{F^c_n(z, \epsilon)} \, \textup{d}\Pi_0(z) \textup{d}P_{\beta_0, z_0}(y) \\
	& =   \int_{D(\beta_n, \rho_n)} \int L_n(\beta_n,z;y) 1_{F^c_n(z, \epsilon)} \, \textup{d}P_{\beta_0, z_0}(y) \textup{d}\Pi_0(z).
	\)
	using Fubini's theorem.
	Using the Fatou's lemma with Assumption 3 as a domination condition, we have 
	\(
	&\limsup_{n\to \infty} \int_{D(\beta_n, \rho_n)} \int L_n(\beta_n,z;y)  1_{F^c_n(z, \epsilon)} \, \textup{d}P_{\beta_0, z_0}(y) \textup{d}\Pi_0(z) \\
	& \leq   \int \limsup_{n\to \infty} 1_{D(\beta_n, \rho_n)} \int L_n(\beta_n,z;y)  1_{F^c_n(z, \epsilon)} \, \textup{d}P_{\beta_0, z_0}(y) \textup{d}\Pi_0(z) \\
	& =  \int \limsup_{n\to \infty} 1_{z=\hat{z}(\hat{\beta}_n)} \int L_n(\beta_n,z;y)   1_{F^c_n(z, \epsilon)} \, \textup{d}P_{\beta_0, z_0}(y) \textup{d}\Pi_0(z) \\
	& \stackrel{(a)}{=}0,
	\)
	where (a) is using $F_n^c\{\hat{z}(\hat{\beta}_n), \epsilon\}\to \varnothing$.
	Hence, 
	$$
	\int_{D(\beta_n, \rho_n)} L_n(\beta_n,z;y) \, \textup{d}\Pi_0(z) 
	= \int_{D(\beta_n, \rho_n)} L_n(\beta_n,z;y) 1_{F_n(z, \epsilon)} \, \textup{d}\Pi_0(z) + o_{P_{\beta_0,z_0}}(1).
	$$
	Let $A_n(z,\epsilon)=\{|r_n(h_n,z)| \leq \epsilon\}$. Then
	\(
	& \int_{D(\beta_n, \rho_n)} L_n(\beta_n,z;y) 1_{F_n(z, \epsilon)}\, \textup{d}\Pi_0(z) \\
	& = \int_{D(\beta_n, \rho_n)} L_n(\beta_n,z;y) 1_{F_n(z, \epsilon)} 1_{A_n(z, \epsilon)} \, \textup{d}\Pi_0(z) + \int_{D(\beta_n, \rho_n)} L_n(\beta_n,z;y) 1_{F_n(z, \epsilon)} 1_{A^c_n(z, \epsilon)} \, \textup{d}\Pi_0(z)\,.
	\)
	Similarly, the integral of the second term
	\(
	&  \int \int_{D(\beta_n, \rho_n)} L_n(\beta_n,z;y) 1_{F_n(z, \epsilon)} 1_{A^c_n(z, \epsilon)} \, \textup{d}\Pi_0(z) \textup{d}P_{\beta_0, z_0}(y) \\
	& =   \int_{D(\beta_n, \rho_n)} \int L_n(\beta_n,z;y) 1_{F_n(z, \epsilon)} 1_{A^c_n(z, \epsilon)} \, \textup{d}P_{\beta_0, z_0}(y) \textup{d}\Pi_0(z).
	\)
	using Fubini's theorem.
	Using the Fatou's lemma with Assumption 3 as a domination condition, we have 
	\(
	&\limsup_{n\to \infty} \int_{D(\beta_n, \rho_n)} \int L_n(\beta_n,z;y)  1_{F_n(z, \epsilon)} 1_{A^c_n(z, \epsilon)} \, \textup{d}P_{\beta_0, z_0}(y) \textup{d}\Pi_0(z) \\
	& \leq   \int \limsup_{n\to \infty} 1_{D(\beta_n, \rho_n)} \int L_n(\beta_n,z;y)  1_{F_n(z, \epsilon)} 1_{A^c_n(z, \epsilon)} \, \textup{d}P_{\beta_0, z_0}(y) \textup{d}\Pi_0(z) \\
	& \stackrel{(a)}{\leq}   \int \limsup_{n\to \infty} \int L_n(\beta_n,z;y)   1_{A^c_n(z, \epsilon)} \, \textup{d}P_{\beta_0, z_0}(y) \textup{d}\Pi_0(z) \\
	& \stackrel{(b)}{=}0,
	\)
	where (a) is using $1_{F_n(z, \epsilon)}\leq 1$ and $1_{D(\beta_n, \rho_n)}\leq 1$, and (b) is using $r_n(h_n,z) = o_{P_{\beta_0,z_0}}(1)$.
	Hence, 
	\[\label{eq:proof1}
	\int_{D(\beta_n, \rho_n)} L_n(\beta_n,z;y) \, \textup{d}\Pi_0(z) 
	= \int_{D(\beta_n, \rho_n)} L_n(\beta_n,z;y) 1_{F_n(z, \epsilon)} 1_{A_n(z, \epsilon)} \, \textup{d}\Pi_0(z) + o_{P_{\beta_0,z_0}}(1).
	\]
	Let $N$ large enough such that $\rho_n < \delta$ for all $n>N$. For all $z$ satisfying $d_H\{z,  \hat{z}(\beta_n); \beta_n\} < \rho_n$, we have $l_n(\beta_n, z) = l_n\{\hat{\beta}_n, \hat{z}(\hat{\beta}_n)\} - \frac{1}{2}  h_n^\T H_n(z) h_n +  r_n(h_n,z)$ by Assumption 2.  Therefore,
	\(
	& \int_{D(\beta_n, \rho_n)} L_n(\beta_n,z;y) 1_{F_n(z, \epsilon)} 1_{A_n(z, \epsilon)} \, \textup{d}\Pi_0(z) \\
	& = \int_{D(\beta_n, \rho_n)} \exp\biggl\{ l_n\{\hat{\beta}_n, \hat{z}(\hat{\beta}_n)\} - \frac{1}{2}  h_n^\T H_n(z) h_n +  r_n(h_n,z) \biggr\} 1_{F_n(z, \epsilon)} 1_{A_n(z, \epsilon)} \, \textup{d}\Pi_0(z).
	\)
	For all $y \in F_n(z, \epsilon)\cap A_n(z, \epsilon)$, we have
	$$
	- \frac{1}{2}  h_n^\T H_0 h_n - 2\epsilon \leq - \frac{1}{2}  h_n^\T H_n(z) h_n +  r_n(h_n,z) \leq - \frac{1}{2}  h_n^\T H_0 h_n + 2\epsilon,
	$$
	so we have the lower and upper bounds
	\(
	& \int_{D(\beta_n, \rho_n)} \exp\biggl\{l_n\{\hat{\beta}_n, \hat{z}(\hat{\beta}_n)\} - \frac{1}{2}  h_n^\T H_0 h_n - 2\epsilon \biggr\} 1_{F_n(z, \epsilon)} 1_{A_n(z, \epsilon)} \, \textup{d}\Pi_0(z) \\
	& \leq  \int_{D(\beta_n, \rho_n)} L_n(\beta_n,z;y) 1_{F_n(z, \epsilon)} 1_{A_n(z, \epsilon)} \, \textup{d}\Pi_0(z) \\
	& \leq \int_{D(\beta_n, \rho_n)} \exp\biggl\{l_n\{\hat{\beta}_n, \hat{z}(\hat{\beta}_n)\} - \frac{1}{2}  h_n^\T H_0 h_n + 2\epsilon \biggr\} 1_{F_n(z, \epsilon)} 1_{A_n(z, \epsilon)} \, \textup{d}\Pi_0(z).
	\)
	
	Especially, when $h_n=0$ and $\beta_n = \hat{\beta}_n$, we have $l_n(\hat{\beta}_n, z) = l_n[\hat{\beta}_n, \hat{z}(\hat{\beta}_n)] +  r_n(h_n,z)$. Hence,
	\(
	& \int_{D(\beta_n, \rho_n)} \exp(l_n\{\hat{\beta}_n, \hat{z}(\hat{\beta}_n)\}- \epsilon ) 1_{F_n(z, \epsilon)} 1_{A_n(z, \epsilon)} \, \textup{d}\Pi_0(z) \\
	& \leq  \int_{D(\beta_n, \rho_n)} {L_n(\hat{\beta}_n,z;y)} 1_{F_n(z, \epsilon)} 1_{A_n(z, \epsilon)} \, \textup{d}\Pi_0(z) \\
	& \leq \int_{D(\beta_n, \rho_n)} \exp(l_n\{\hat{\beta}_n, \hat{z}(\hat{\beta}_n)\}+ \epsilon ) 1_{F_n(z, \epsilon)} 1_{A_n(z, \epsilon)} \, \textup{d}\Pi_0(z).
	\)
	We combine the above two group of inequalities to induce
	\(
	& \exp\biggl\{ - \frac{1}{2}  h_n^\T H_0 h_n - 3\epsilon \biggr\} \int_{D(\beta_n, \rho_n)} {L_n(\hat{\beta}_n,z;y)}  1_{F_n(z, \epsilon)} 1_{A_n(z, \epsilon)} \, \textup{d}\Pi_0(z) \\
	& \leq  \int_{D(\beta_n, \rho_n)} L_n(\beta_n,z;y) 1_{F_n(z, \epsilon)} 1_{A_n(z, \epsilon)} \, \textup{d}\Pi_0(z) \\
	& \leq \exp\biggl\{ - \frac{1}{2}  h_n^\T H_0 h_n + 3\epsilon \biggr\} \int_{D(\beta_n, \rho_n)} {L_n(\hat{\beta}_n,z;y)} 1_{F_n(z, \epsilon)} 1_{A_n(z, \epsilon)} \, \textup{d}\Pi_0(z).
	\)
	Hence,
	\(
	&  - \frac{1}{2}  h_n^\T H_0 h_n - 3\epsilon +\log \int_{D(\beta_n, \rho_n)} L_n(\hat{\beta}_n,z;y)  1_{F_n(z, \epsilon)} 1_{A_n(z, \epsilon)} \, \textup{d}\Pi_0(z) \\
	& \leq  \log\int_{D(\beta_n, \rho_n)} L_n(\beta_n,z;y) 1_{F_n(z, \epsilon)} 1_{A_n(z, \epsilon)} \, \textup{d}\Pi_0(z) \\
	& \leq  - \frac{1}{2}  h_n^\T H_0 h_n + 3\epsilon +\log \int_{D(\beta_n, \rho_n)} L_n(\hat{\beta}_n,z;y)  1_{F_n(z, \epsilon)} 1_{A_n(z, \epsilon)} \, \textup{d}\Pi_0(z).
	\)
	By \eqref{eq:proof1}, we have
	\(
	&  - \frac{1}{2}  h_n^\T H_0 h_n - 3\epsilon +\log \int_{D(\beta_n, \rho_n)} L_n(\hat{\beta}_n,z;y)   \, \textup{d}\Pi_0(z) + o_{P_{\beta_0,z_0}}(1) \\
	& \leq  \log\int_{D(\beta_n, \rho_n)} L_n(\beta_n,z;y)  \, \textup{d}\Pi_0(z) \\
	& \leq  - \frac{1}{2}  h_n^\T H_0 h_n + 3\epsilon +\log \int_{D(\beta_n, \rho_n)} L_n(\hat{\beta}_n,z;y)   \, \textup{d}\Pi_0(z) + o_{P_{\beta_0,z_0}}(1).
	\)
	Hence, we have
	\(
	\log\int_{D(\beta_n, \rho_n)} L_n(\beta_n,z;y)  \, \textup{d}\Pi_0(z) = \log \int_{D(\beta_n, \rho_n)} L_n(\hat{\beta}_n,z;y) \, \textup{d}\Pi_0(z) - \frac{1}{2}  h_n^\T H_0 h_n + o_{P_{\beta_0,z_0}}(1).
	\)
	This suffices to prove Lemma 1.
\end{proof}

\begin{proof}[of Theorem 2]
	We have $\pi_n(\beta) = \exp\{{s}_n(\beta)\} \pi_0(\beta) / m_n$ \\where $m_n = \int_{\mathbb{R}^d} \exp\{ s_n(\beta)\}\pi_0(\beta) \,\textup{d}\beta$, and
	$q_n(x) = \pi_n(\hat{\beta}_n+x/\sqrt{n}) n^{-d/2}$. Let
	\(g_n(x) =  q_n(x)\exp\{ -s_n(\hat{\beta}_n)\} n^{d/2}m_n = \exp \{s_n(\hat{\beta}_n + x/\sqrt{n}) - s_n(\hat{\beta}_n)\}  \pi_0(\hat{\beta}_n+x/\sqrt{n})
	\)
	and define $g_0(x) = \exp\{ -x^\T H_0 x / 2\}\pi_0(\beta_0)$.
	We first show that
	$\int_{B} |g_n(x) - g_0(x) |\,\textup{d}x \xrightarrow[n\to\infty]{P_{\beta_0,z_0}} 0$, where $B = \{x: \|x\| \leq M\}$ for any fix positive number $M$.
	Since $\pi_0$ is continuous at $\beta_0$, we choose sufficiently small $\epsilon$ such that $\pi_0(\beta) \leq 2\pi_0(\beta_0) $ for all $\beta\in B_{\epsilon}(\beta_0)$. 
	
	Since $	s_n(\hat{\beta}_n + x/\sqrt{n}) - s_n(\hat{\beta}_n) = -\frac{1}{2} x^\T H_0 x + o_{P_{\beta_0,z_0}}(1)	$,
	and $\pi_0$ is continuous at $\beta_0$ and $\hat{\beta}_n+x/\sqrt{n}\to \beta_0$, we have $g_n(x)\to g_0(x)$ pointwise with probability converge to $1$.
	Consider $n$ sufficiently large such that the term $o_{P_{\beta_0,z_0}}(1) < \alpha / 2$ where $\alpha$ is less than the smallest eigenvalue of $H_0$. We denote $A_0 = H_0 - \alpha I$, and define
	\(
	h_0(x) = \exp( - x^\T A_0 x /2 ) 2\pi_0(\beta_0).
	\)
	When $\|x\| \leq  M$, for $n$ large enough we have $\|(\hat{\beta}_n+x/\sqrt{n}) - \beta_0\| < \epsilon$. By the choice of $\epsilon$, we have $\pi_0(\hat{\beta}_n+x/\sqrt{n}) \leq 2\pi_0(\beta_0)$. Hence $g_n(x) \leq h_0(x)$ with probability converge to $1$ for $n$ sufficiently large.
	Since $g_n,g_0,h_0$ are integrable, by the dominated convergence theorem (the version for convergence in probability), we have $\int_{B} |g_n(x) - g_0(x) |\,\textup{d}x \xrightarrow[n\to\infty]{P_{\beta_0,z_0}} 0$ and $\int_{B} g_n(x) \,\textup{d}x \xrightarrow[n\to\infty]{P_{\beta_0,z_0}} \int_{B} g_0(x) \,\textup{d}x$.
	
	Let $a_n = 1/\int_{B} g_n(x)\,\textup{d}x$ and $a_0 = 1/\int_{B} g_0(x)\,\textup{d}x $. Then $a_n\to a_0$ in $P_{\beta_0,z_0}$-probability, and thus
	\(
	&\int_{B} \bigg|\frac{q_n(x)}{\int_B q_n} - \mathcal{N}^B\bigl(x \mid 0, H_{0}^{-1}\bigr) \bigg|\,\textup{d}x
	= \int_{B} |a_n g_n(x) - a_0 g_0(x) |\,\textup{d}x \\
	& \leq \int_{B} |a_n g_n(x) - a_n g_0(x)|\,\textup{d}x + \int_{B} |a_n g_0(x) - a_0 g_0(x) |\,\textup{d}x\\
	&  \leq |a_n| \int_{B} |g_n(x)-g_0(x)|\,\textup{d}x + |a_n-a_0| \int_{B} |g_0(x)|\,\textup{d}x \xrightarrow[n\to\infty]{P_{\beta_0,z_0}} 0.
	\)
	
	Then we can choose slowly enough increasing $M_n\to \infty$ such that 
	$$
	\int_{B_n} \bigg|\frac{q_n(x)}{\int_{B_n} q_n} - \mathcal{N}^{B_n}\bigl(x \mid 0, H_0^{-1}\bigr) \bigg| \, dx \xrightarrow[n\to\infty]{P_{\beta_0,z_0}} 0
	$$
	where $B_n=\{x: \|x\|\leq M_n\}$.	
	By \citet[Lemma 6.1]{bickel2012semiparametric}, $\Pi(B_n^c) = \int_{B_n^c} q_n(x)\,dx \xrightarrow[n\to\infty]{P_{\beta_0,z_0}} 0$. 	Since $\|\Pi - \Pi^B\| \leq 2\Pi(B^c)$, this proves $\text{d}_{\text{TV}}\bigl\{q_n,\mathcal{N}\bigl(0,H_0^{-1}\bigr)\bigr\} \xrightarrow[n\to\infty]{P_{\beta_0,z_0}} 0$, and the concentration follows from \citet[Lemma 28]{miller2021asymptotic}.
\end{proof}

\section{Additional Simulation results}

\subsection{Flow network}

\begin{figure}[H]
	
	\vspace{0.3cm}
	
	\centering
	\begin{overpic}[width=.19\textwidth]{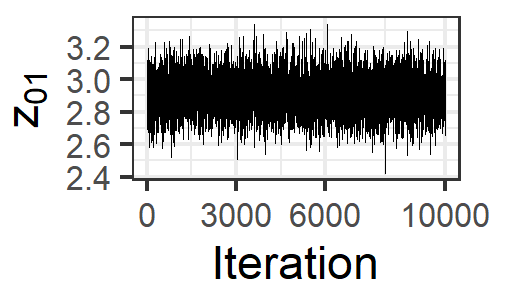}
		\put(24,62){\small (a) Using our suggested inverse mass matrix}
	\end{overpic}
	\includegraphics[width=.19\textwidth]{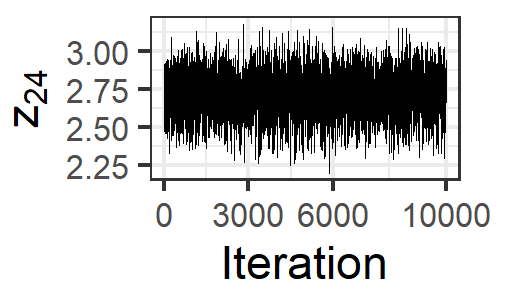}
	\includegraphics[width=.19\textwidth]{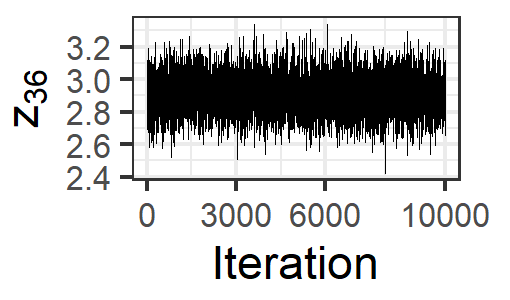}
	\includegraphics[width=.19\textwidth]{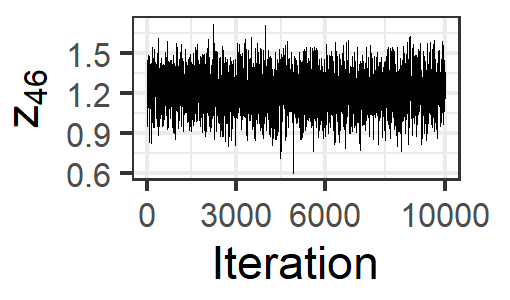}
	\includegraphics[width=.19\textwidth]{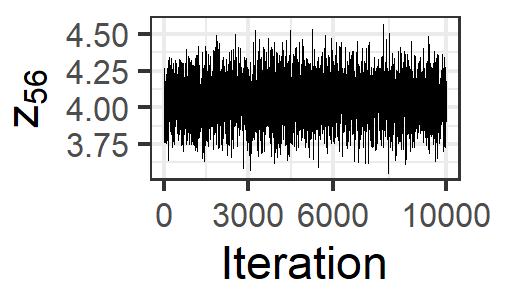}
	\includegraphics[width=.19\textwidth]{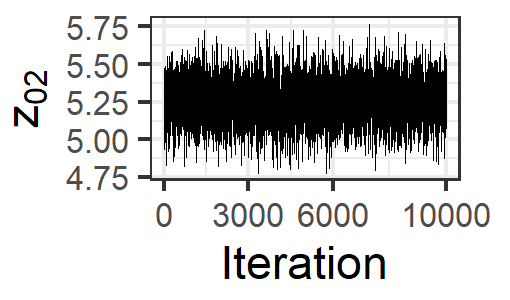}
	\includegraphics[width=.19\textwidth]{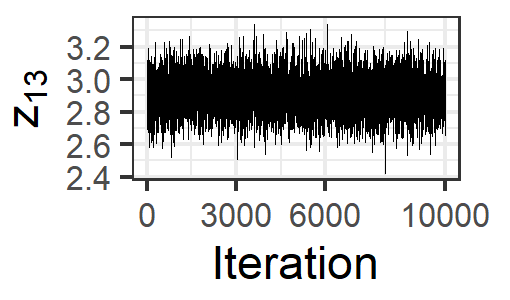}
	\includegraphics[width=.19\textwidth]{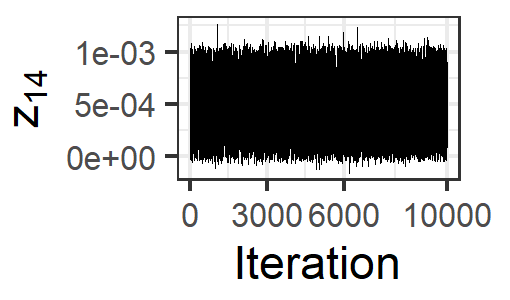}
	\includegraphics[width=.19\textwidth]{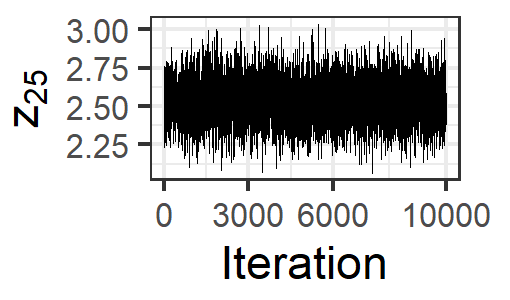}
	\includegraphics[width=.19\textwidth]{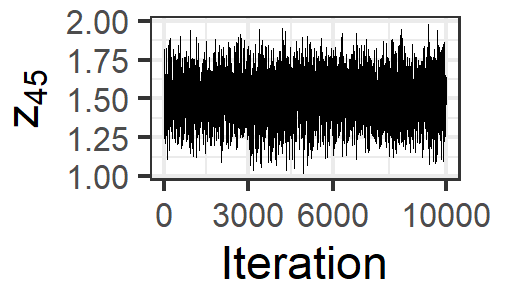}
	
	\vspace{0.6cm}
	
	\centering
	\begin{overpic}[width=.19\textwidth]{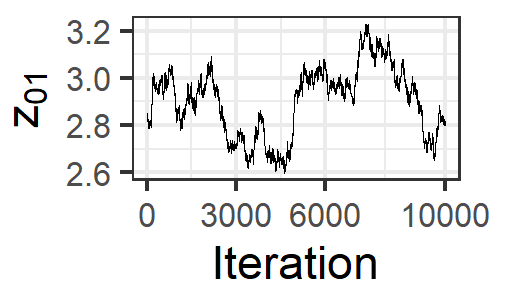}
		\put(24,62){\small (b) Using the default adapted matrix as the inverse mass}
	\end{overpic}
	\includegraphics[width=.19\textwidth]{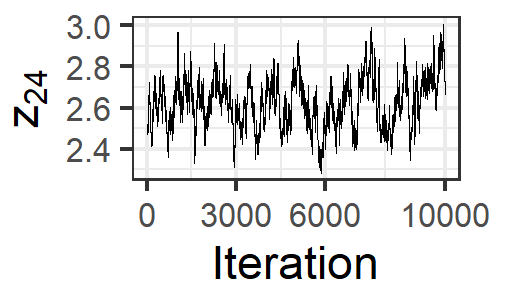}
	\includegraphics[width=.19\textwidth]{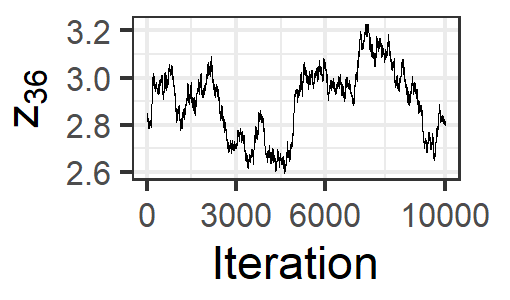}
	\includegraphics[width=.19\textwidth]{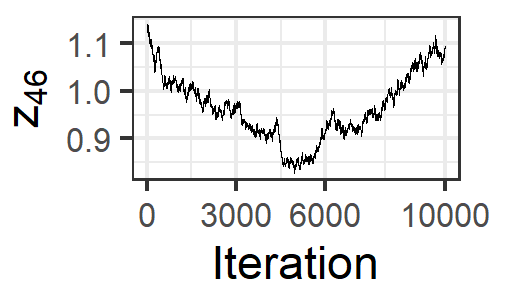}
	\includegraphics[width=.19\textwidth]{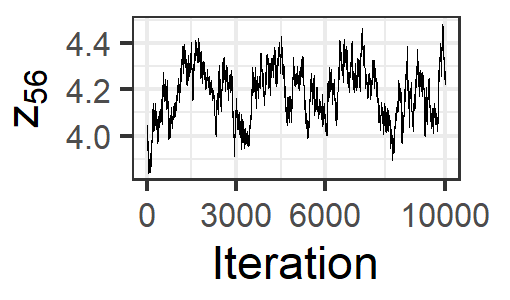}
	\includegraphics[width=.19\textwidth]{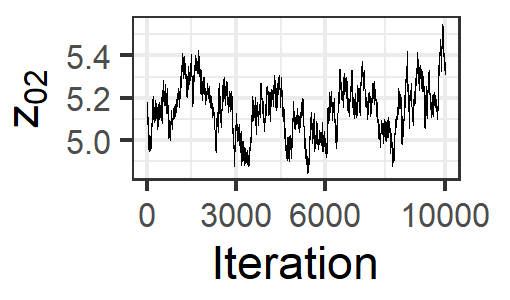}
	\includegraphics[width=.19\textwidth]{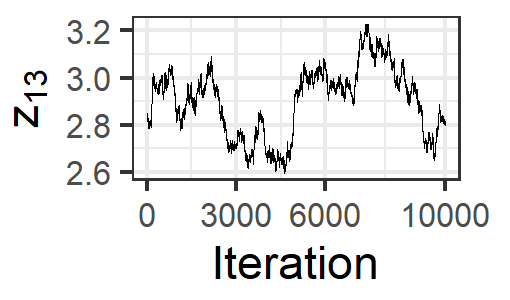}
	\includegraphics[width=.19\textwidth]{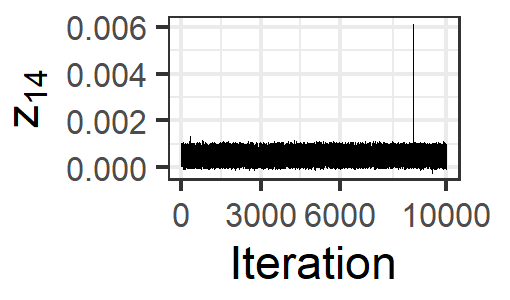}
	\includegraphics[width=.19\textwidth]{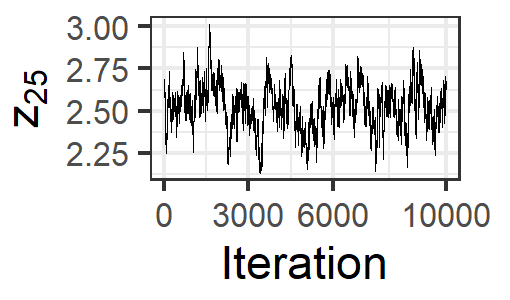}
	\includegraphics[width=.19\textwidth]{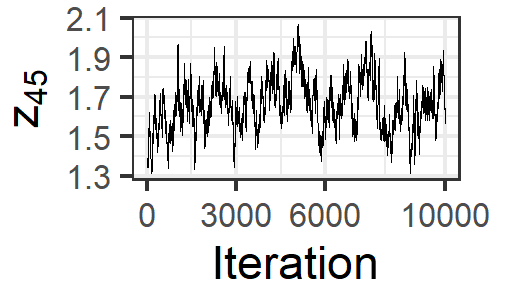}
	
	\caption{Traceplots of Markov chain for sampling the posterior of $z$ using different choices of inverse mass matrix.}
	\label{fig:combined-traces}
\end{figure}

\begin{figure}[H]
	
	\vspace{0.3cm}
	
	\centering
	\begin{overpic}[width=.19\textwidth]{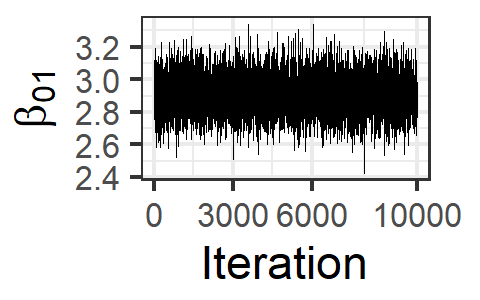}
		\put(26,66){\small (a) Using our suggested inverse mass matrix}
	\end{overpic}
	\includegraphics[width=.19\textwidth]{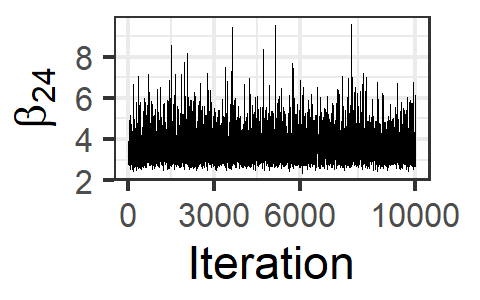}
	\includegraphics[width=.19\textwidth]{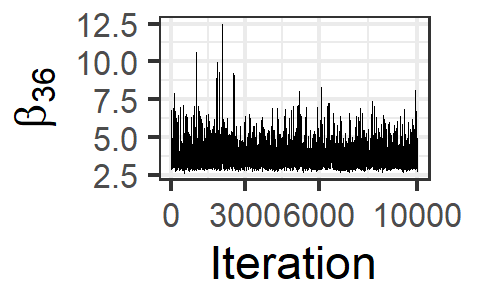}
	\includegraphics[width=.19\textwidth]{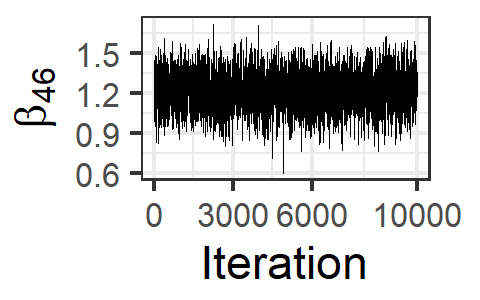}
	\includegraphics[width=.19\textwidth]{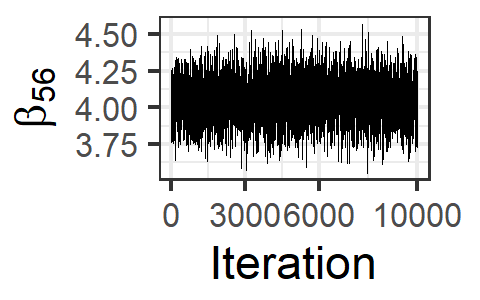}
	\includegraphics[width=.19\textwidth]{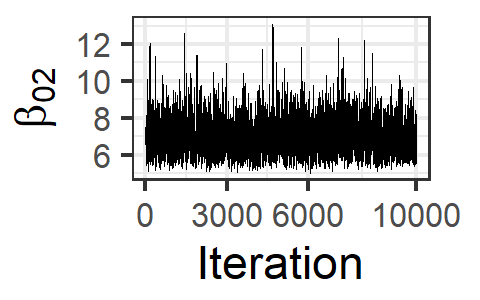}
	\includegraphics[width=.19\textwidth]{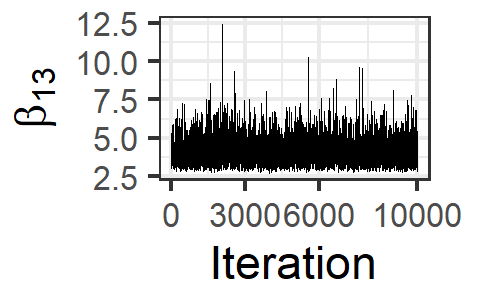}
	\includegraphics[width=.19\textwidth]{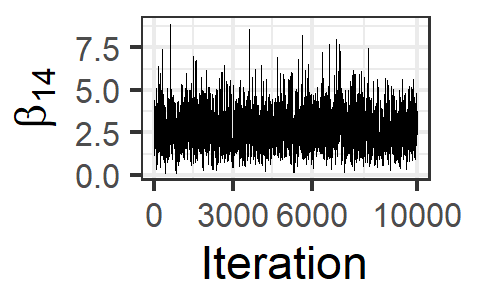}
	\includegraphics[width=.19\textwidth]{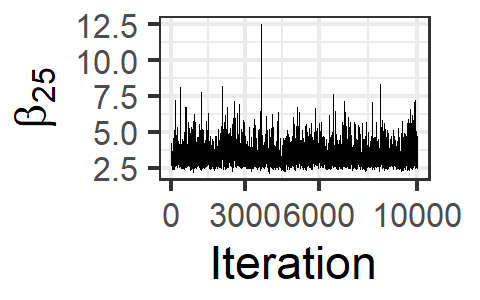}
	\includegraphics[width=.19\textwidth]{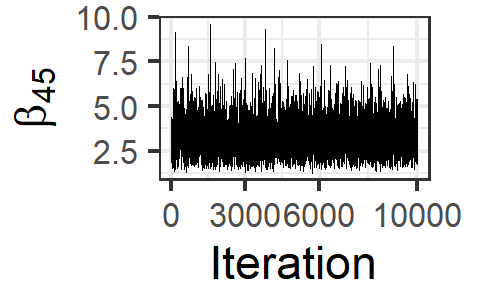}
	
	\vspace{0.6cm}
	
	\centering
	\begin{overpic}[width=.19\textwidth]{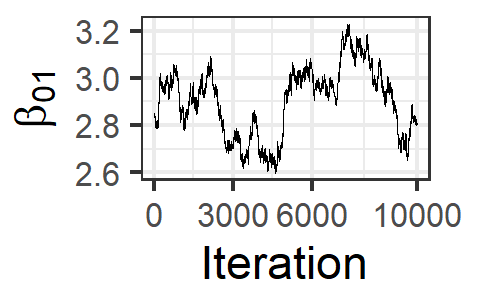}
		\put(26,66){\small (b) Using the default adapted matrix as the inverse mass}
	\end{overpic}
	\includegraphics[width=.19\textwidth]{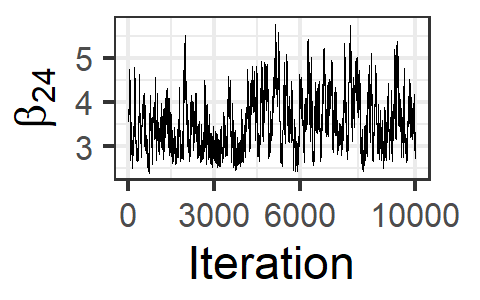}
	\includegraphics[width=.19\textwidth]{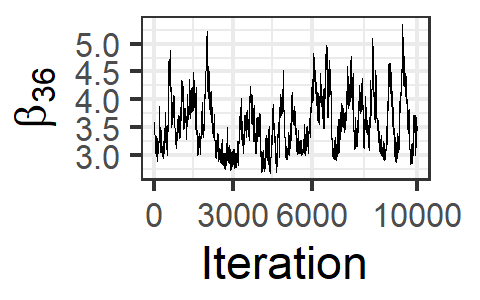}
	\includegraphics[width=.19\textwidth]{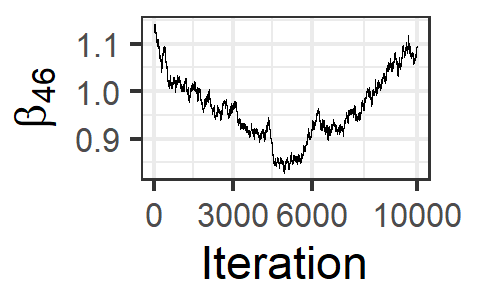}
	\includegraphics[width=.19\textwidth]{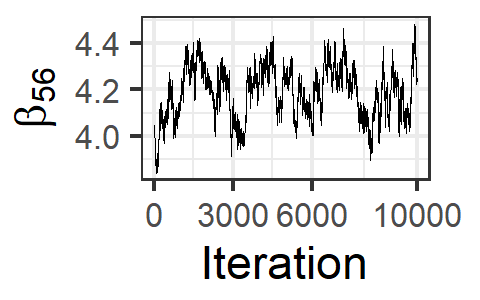}
	\includegraphics[width=.19\textwidth]{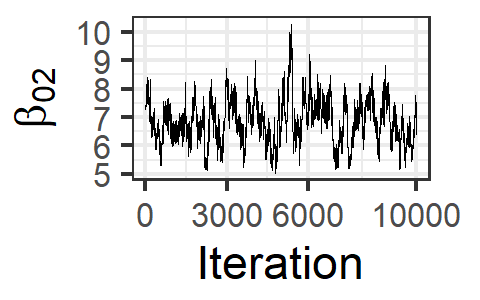}
	\includegraphics[width=.19\textwidth]{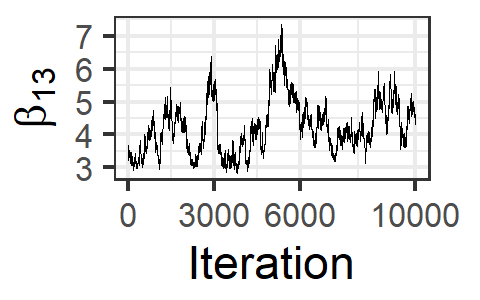}
	\includegraphics[width=.19\textwidth]{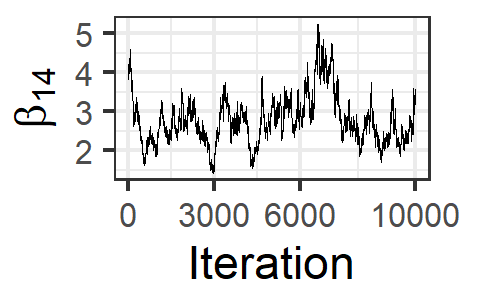}
	\includegraphics[width=.19\textwidth]{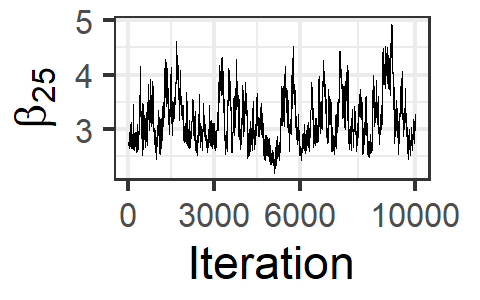}
	\includegraphics[width=.19\textwidth]{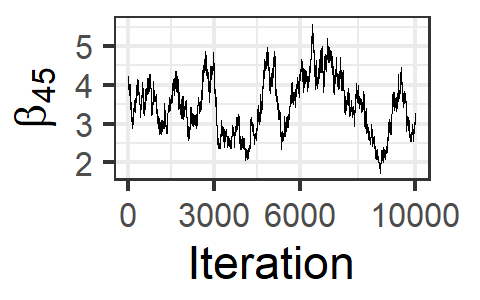}
	
	\caption{Traceplots of Markov chains for sampling the posterior of $\beta$ using different choices of inverse mass matrix.}    \label{fig:combined-beta-traces}
\end{figure}

Figure \ref{fig:combined-traces} and \ref{fig:combined-beta-traces} show the traces of the Markov chains of the posterior samples of $z$ and $\beta$ in Section 5.

\subsection{Latent quadratic model}

We use another numerical experiment on latent quadratic model to illustrate the advantage of the dual form. Consider the canonical latent normal model with the likelihood kernel:
\(
g(y;\beta, z) = \exp\biggl\{ -\frac{1}{2} z^\T Q^{-1}(\beta;x) z \biggr\} \prod_{i=1}^n v(y_i \mid z_i),
\)
where $v(y_i \mid z_i)$ is a log-concave conditional density of $y_i$, and $y_1,\dots,y_n$ are assumed to be conditionally independent given $z$. The covariance structure of the latent variable $z$ is parameterized by $Q(\beta;x)$, where $Q(\beta;x)_{i,j}= \tau \exp\{- \|x_i-x_j\|^2/(2b)\}$ with $x_i \in \mathbb{R}^{d}$ representing observed predictors or spatial locations. The parameter $\beta=(\tau,b)\in \mathbb{R}^2$ controls the scale of dependence in the covariance structure. As a concrete example, we consider binary observations $y_i$ from Bernoulli distribution under logistic link $v(y_i \mid z_i) = \exp(y_i z_i) / \{1+\exp(z_i)\}$. This formulation corresponds to a latent Gaussian process model for binary classification, where $z$ acts as an underlying continuous latent variable governing the probability of success.

To create a gradient-bridged posterior, we seek to minimize $-\log g(y;\beta, z)$ over $z\in \mathbb{R}^n$. We define the optimization function $h(\beta,z;y) = -\log g(y;\beta, z)$. However, direct posterior computation involves the inversion $Q^{-1}(\beta;x)$, which is computationally expensive with a complexity of $O(n^3)$. Hence, we use the dual form to address this inefficiency.

Since $-\log g(y;\beta, z)$ can be decomposed into the sum of a quadratic function and a convex function, we use the variable splitting technique by introducing the constraint $u=z$. Introducing the Lagrange multiplier $\alpha\in \mathbb{R}^{n}$, the Lagrangian dual function takes the form:
\(
h^\dagger (\beta,\alpha;y) & = \inf_{z,u}\ \frac{1}{2} z^\T Q^{-1}(\beta;x) z + \alpha^\T (z-u) + \sum_{i=1}^n \bigl[ -y_i u_i  + \log\{1+ \exp(u_i)\} \bigr]\\
& = -\frac{1}{2} \alpha^\T Q(\beta;x) \alpha   - \sum_{i=1}^n \bigl\{ (\alpha_i+y_i) \log(\alpha_i+y_i)  + (1-\alpha_i-y_i) \log(1-\alpha_i-y_i) \bigr\}
\)
subject to $\alpha_i+y_i\in (0,1)\ (i=1,\dots,n)$. Strong duality holds in this setting, ensuring that $\sup_\alpha h^\dagger (\beta, \alpha;y) = \inf_z h(\beta,z;y)$, where the optimal solutions satisfy $\hat{z} = -Q(\beta;x) \hat \alpha$ with $\hat{z} = \arg\inf_z h(\beta,z;y)$ and $\hat \alpha = \arg\sup_\alpha h^\dagger (\beta, \alpha;y)$. This allows us to reparameterize $z = -Q(\beta;x) \alpha$, and then use $-h^\dagger$ as the gradient-bridge function, leading to the gradient-bridged posterior:
\(
L(y; \beta, \alpha) \propto g\{y; \beta, -Q(\beta;x) \alpha \} \exp \bigl\{ -\lambda \| \nabla_\alpha h^\dagger (\beta,\alpha;y) \|_2^2 \bigr\} \,,
\)
where we can explicitly calculate
\(
\nabla_\alpha h^\dagger (\beta,\alpha;y) = - Q(\beta;x) \alpha - \log(\alpha+y) + \log(1-\alpha-y)\,.
\)
This guarantees that the conditional maximum likelihood estimator is $\alpha=\hat{\alpha}$, which coincides with $\hat z=-Q(\beta;x) \hat{\alpha}$.

A key advantage of this approach is that neither the gradient-bridge function nor its gradient with respect to $\alpha$ requires the inversion $Q^{-1}$, significantly improving the posterior computational efficiency. Additionally, evaluating $g(y;\beta,z)$ remains quick since $z^\T Q^{-1}(\beta;x)z = \alpha^\T  Q(\beta;x) \alpha$.

To simulate data for benchmarking, we generate 1000 random locations $x_1, \dots, x_{1000} \sim \text{Uniform}(-6, 6)$. The ground truth latent curve $\tilde z$ is defined using a cubic spline interpolation with 20 control points, which are evenly spaced along $[-6, 6]$, with their corresponding $z$-values sampled from $\text{Uniform}(-3, 3)$. The spline is evaluated at each $x_i$ to produce $\tilde{z}_i$. Finally, binary observations are generated as $y_i \sim \text{Bernoulli}\bigl[1 / \{1 + \exp(-\tilde{z}_i)\}\bigr]$.

We fit the gradient-bridged posterior, the bridged posterior and the Gibbs posterior to the simulated data. For all models, we assign independent $\text{Ga}^{-1}(2,0.1)$ priors on $\tau$ and $b$. We use no-u-turn sampler for the gradient-bridged posterior and the Gibbs posterior, and use random walk Metropolis for the bridged posterior.

\begin{figure}[ht]
	\centering
	\includegraphics[width=.23\textwidth]{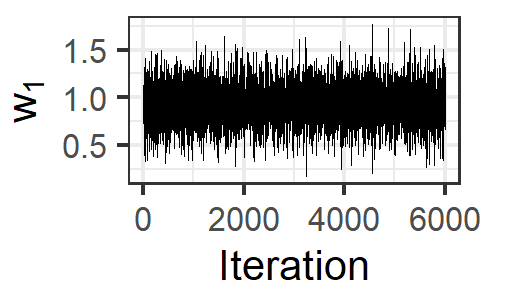}
	\includegraphics[width=.23\textwidth]{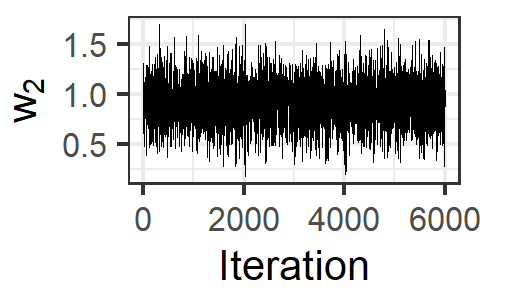}
	\includegraphics[width=.23\textwidth]{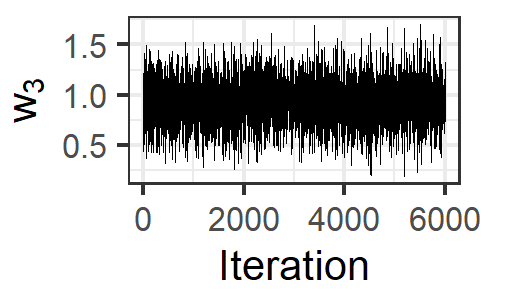}
	\includegraphics[width=.28\textwidth]{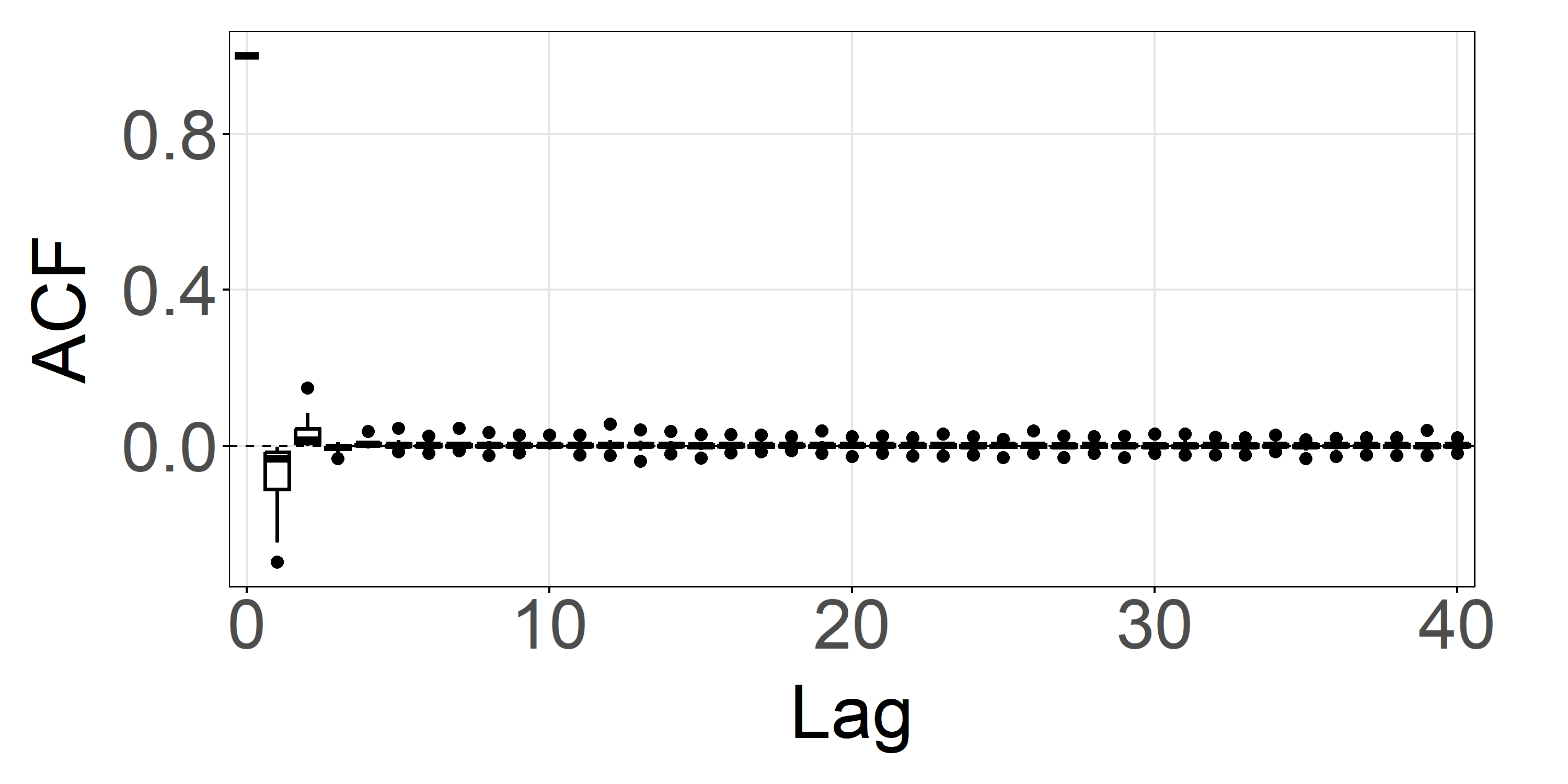}
	\caption{The trace of the Markov chain of posterior samples for the first 3 components of $w$, and the autocorrelation functions for all elements of $w$.}
	\label{fig:w-trace}
\end{figure}

\begin{figure}[ht]
	\centering
	\includegraphics[width=.24\textwidth]{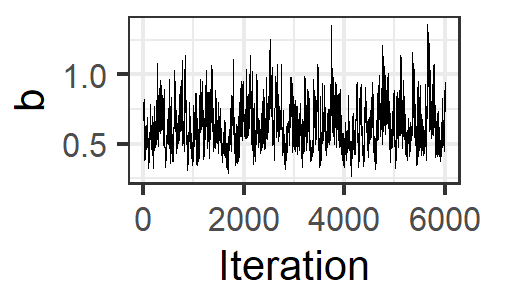}
	\includegraphics[width=.24\textwidth]{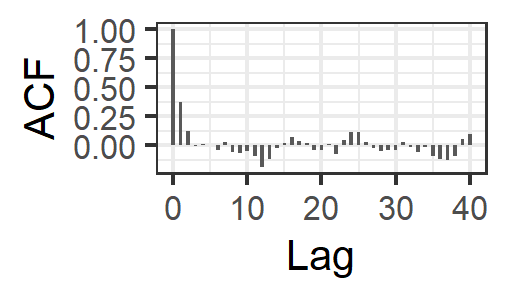}
	\includegraphics[width=.24\textwidth]{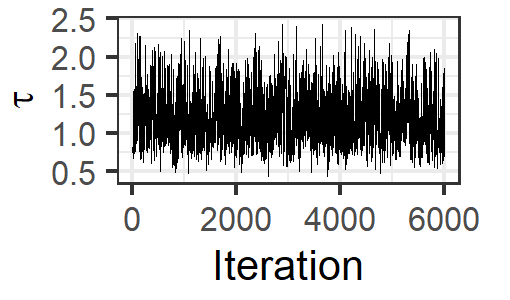}
	\includegraphics[width=.24\textwidth]{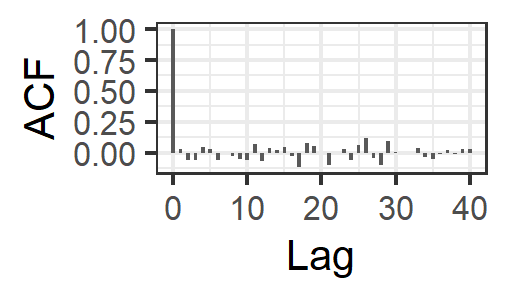}
	\caption{Traces and autocorrelation functions for the posterior samples of $b$ and $\tau$.}
	\label{fig:trace_acf_latent}
\end{figure}

We run each MCMC algorithm for $10,000$ iterations and discard the first $4,000$ as burn-ins, and apply thinning at $20$. 
Figure \ref{fig:w-trace} shows the traces of the first 3 elements of $w$, and the autocorrelation functions for all elements of $w$. The mixing performance is very good. Figure \ref{fig:trace_acf_latent} shows the mixing of the parameters $b$ and $\tau$.

\begin{figure}[H]
	\centering
	\includegraphics[width=.3\textwidth]{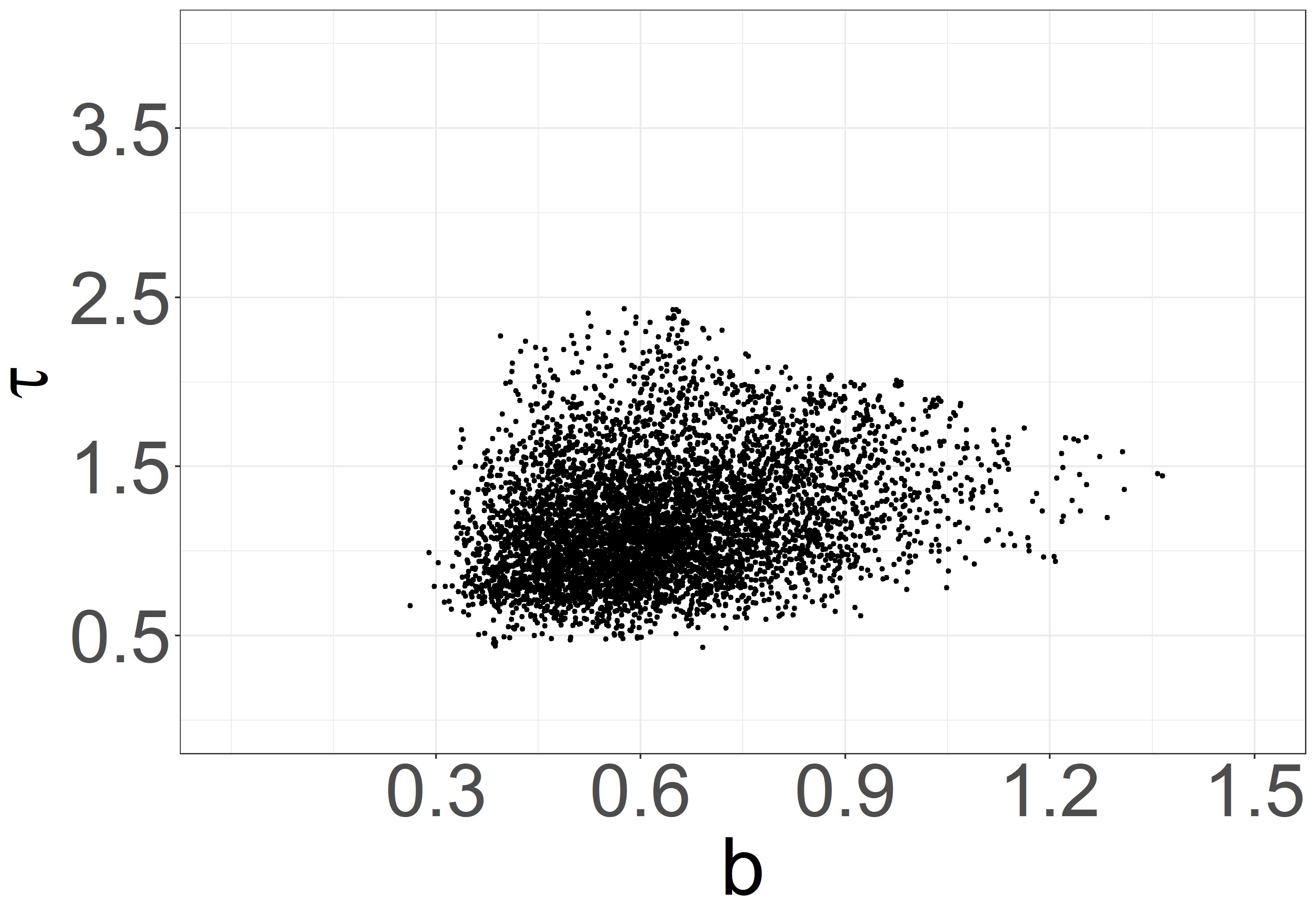}
	\includegraphics[width=.3\textwidth]{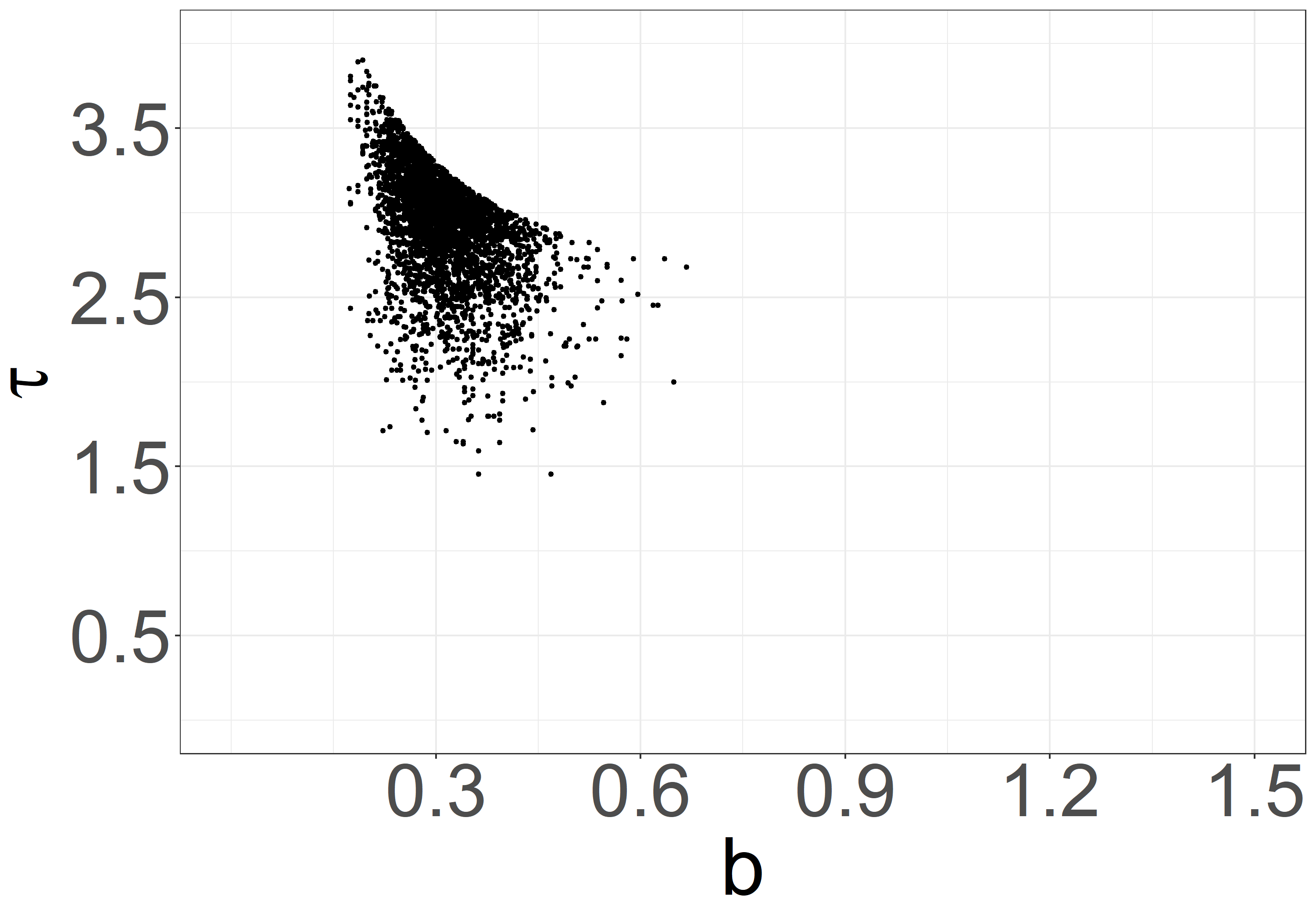}
	\includegraphics[width=.3\textwidth]{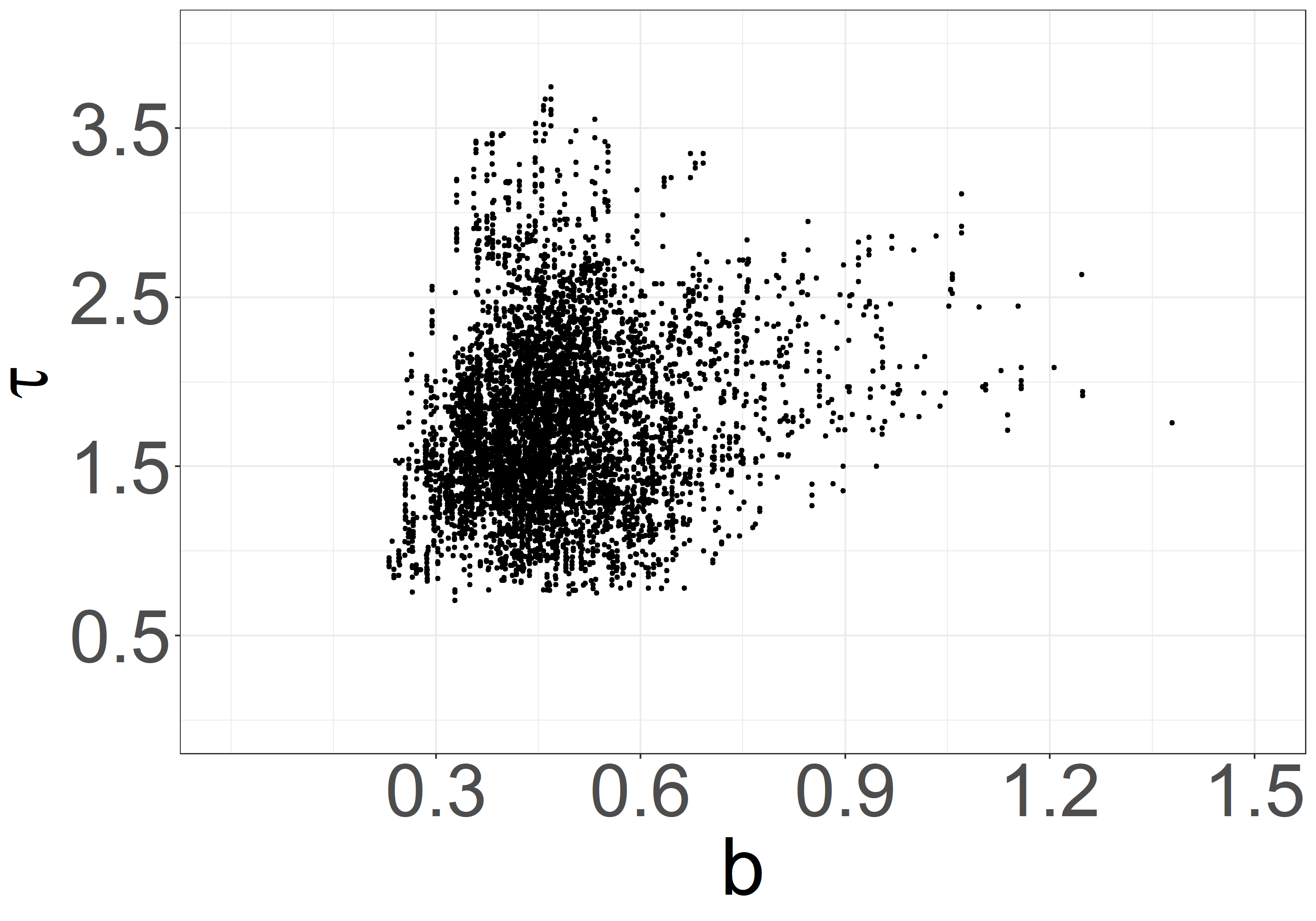}
	\caption{The posterior distributions of the covariance kernel parameters $\tau$ and $b$  for the gradient-bridged posterior (left), the bridged posterior (middle) and the Gibbs posterior (right).}
	\label{fig:latent_gau_post_dist}
\end{figure}

We compare the posterior distributions of parameters $(\tau,b)$. As can be seen in Fig. \ref{fig:latent_gau_post_dist}, these distributions show a similar range of $\tau$ and $b$. Since the distributions of $\tau$ or $b$ arise from different models, their distributions are not expected to align precisely.

\section{Data preprocessing for single-cell data integration}

The preprocessing of the panc8 dataset is conducted by first grouping the data based on sequencing technology. Within each group, gene expression levels are standardized to ensure comparability. Then, highly variable genes are identified to capture the most informative features, followed by the application of principal component analysis. The number of principal components to retain is determined using a scree plot, with five components selected as the optimal choice.

To standardize sample sizes across all technology groups, the group with the smallest number of samples is identified, and denoted as $X_{\text{raw}}^*$. Cell types within $ X_{\text{raw}}^* $ containing  20 or fewer samples are removed to prevent the influence of small sample sizes on integration. For the remaining groups, downsampling is performed using stratified sampling to ensure that the sample counts for each cell type matched those in $X_{\text{raw}}^*$. If the available sample count for a given cell type in a group is insufficient to meet the target, 
additional samples are randomly drawn from other cell types within the group to maintain the overall sample size. This procedure ensured consistency in sample sizes across all groups while preserving the distribution of cell types.

\section{Mixing performance for data integration application}

\begin{figure}[ht]
	
	\vspace{0.5cm}
	
	\centering
	\begin{overpic}[width=0.41\textwidth]{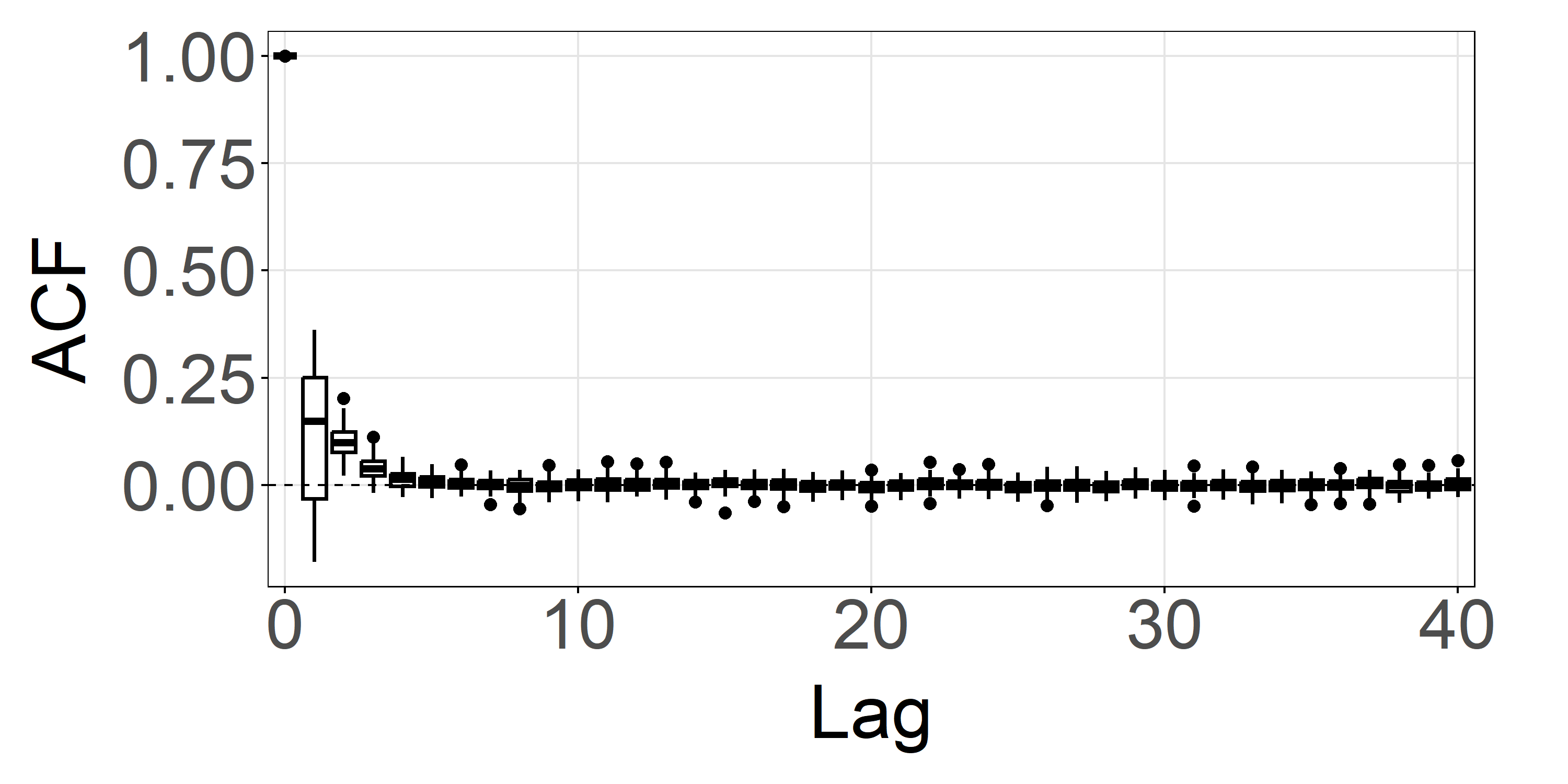}
		\put(17,51){\small (a) $R_{ij}$'s, gradient-bridged posterior}
	\end{overpic}
	\begin{overpic}[width=0.41\textwidth]{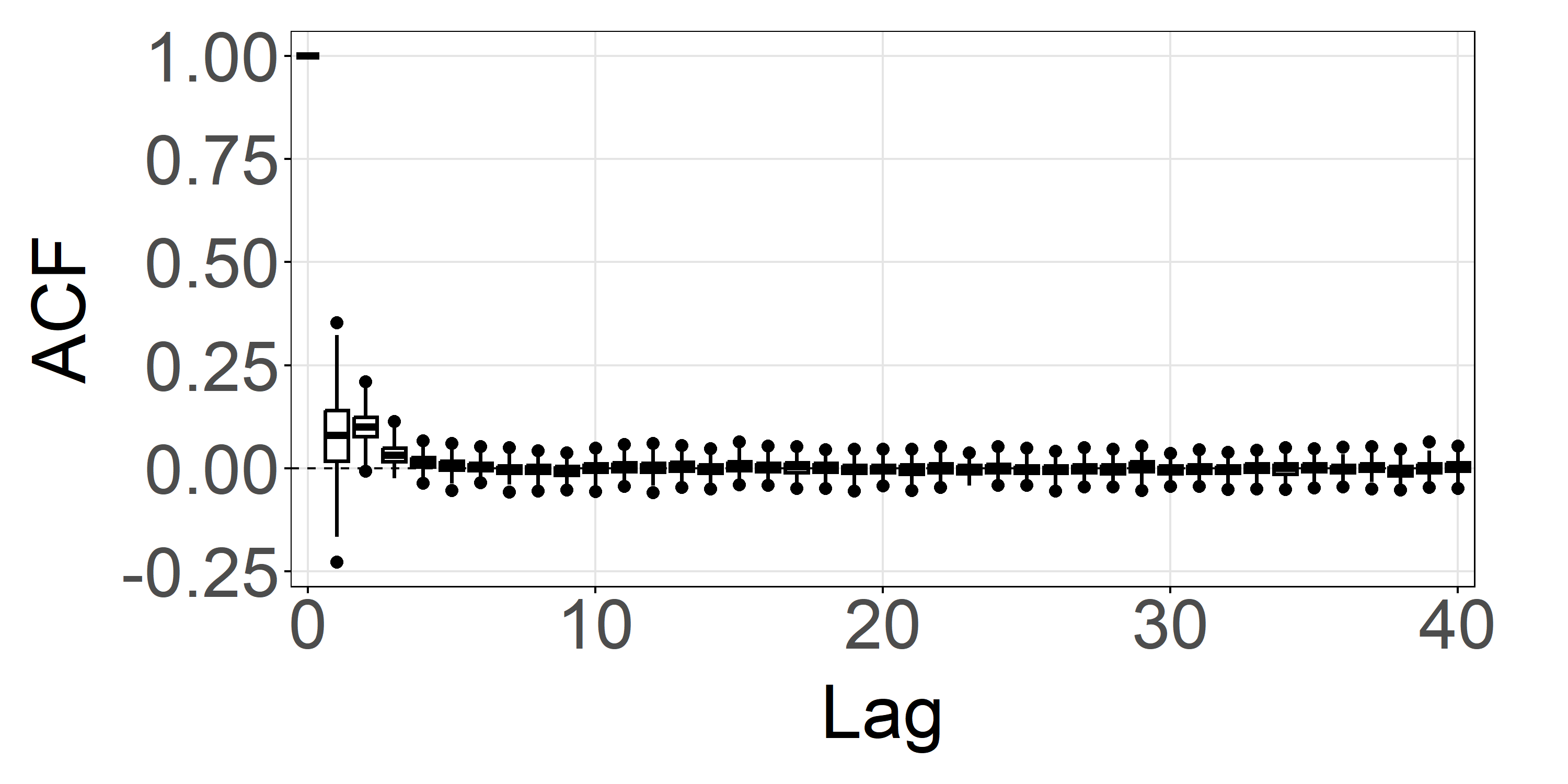}
		\put(17,51){\small (b) $u_{ij}$'s, gradient-bridged posterior}
	\end{overpic}
	
	\vspace{0.5cm}
	
	\centering
	\begin{overpic}[width=0.41\textwidth]{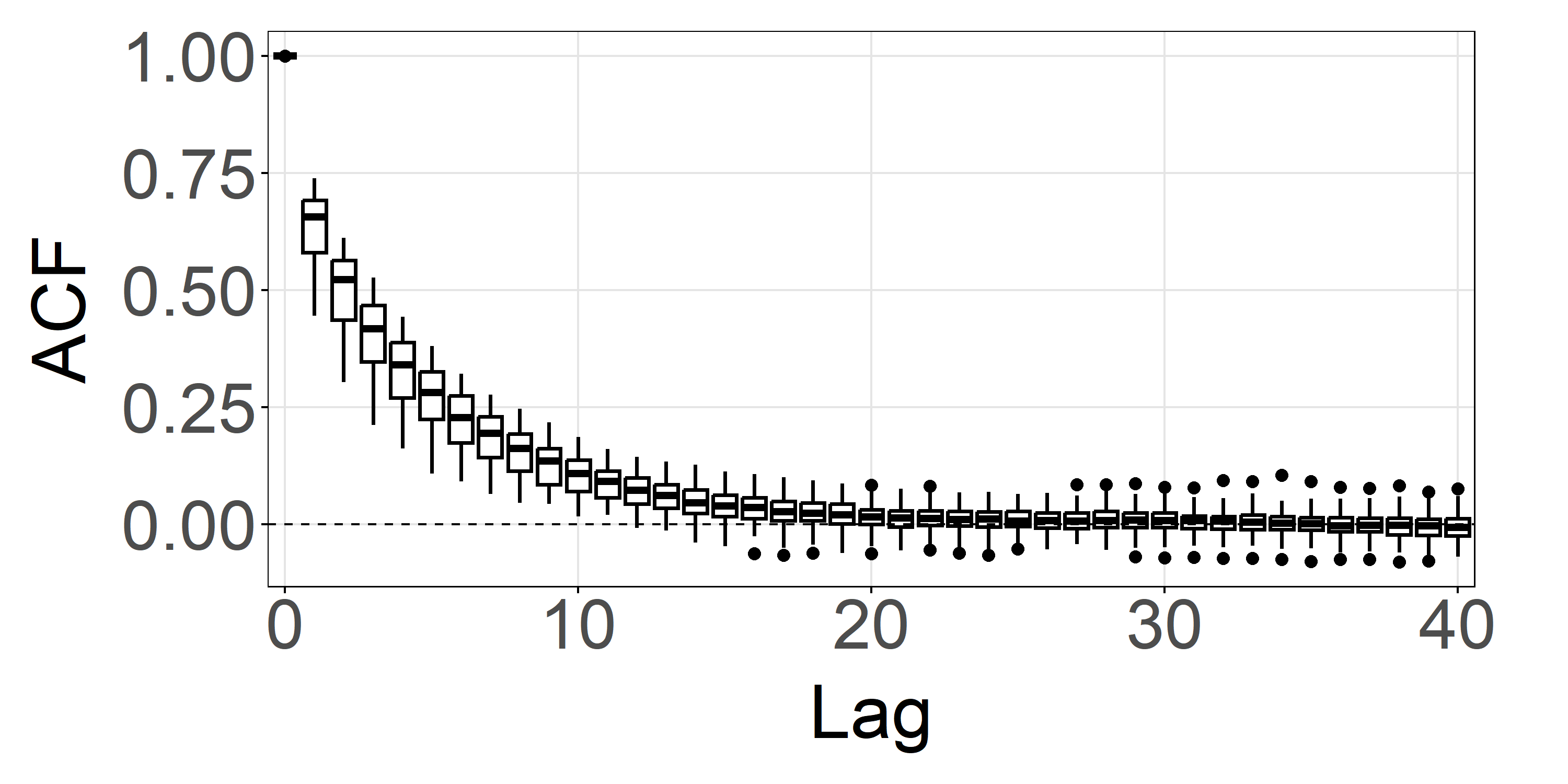}
		\put(17,51){\small (c) $R_{ij}$'s, Gibbs posterior}
	\end{overpic}
	\begin{overpic}[width=0.41\textwidth]{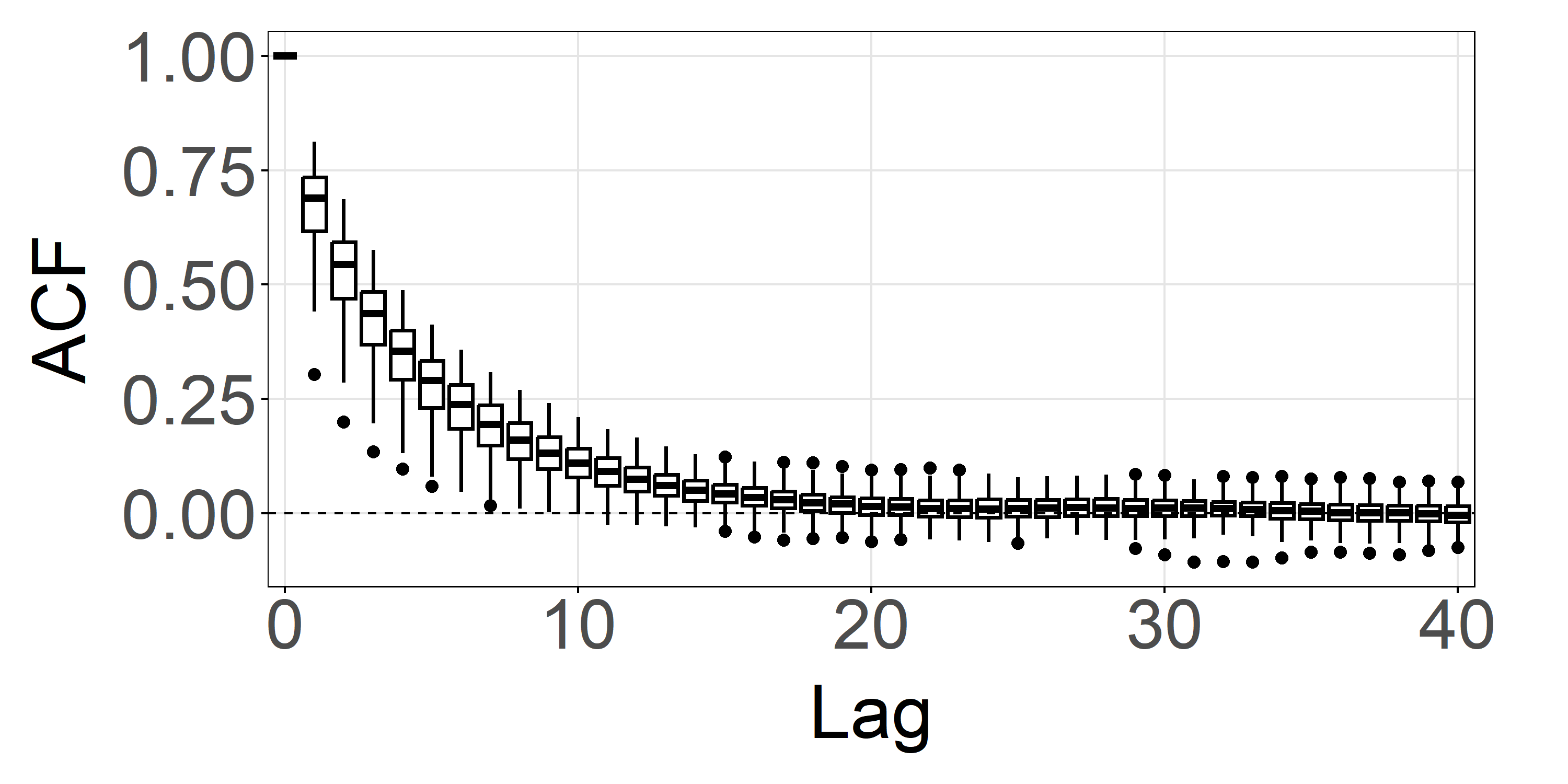}
		\put(17,51){\small (d) $u_{ij}$'s, Gibbs posterior}
	\end{overpic}
	
	\caption{Autocorrelation of the Markov chain samples from different models.}
	\label{fig:data-integration-acf}
\end{figure}

\begin{figure}[ht]
	\centering
	\includegraphics[width=.24\textwidth]{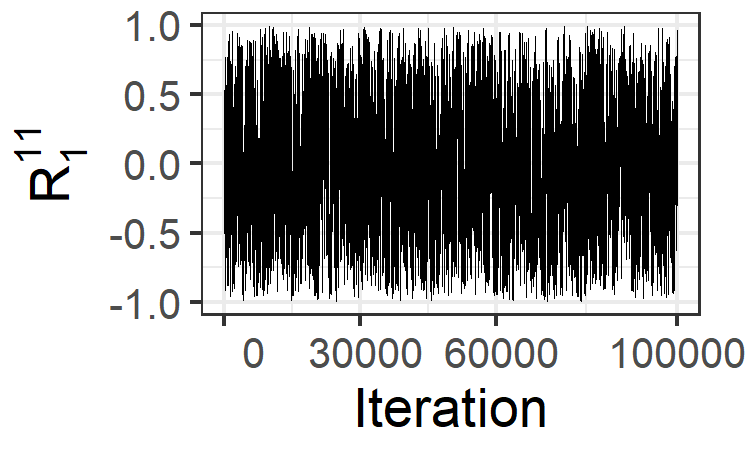}
	\includegraphics[width=.24\textwidth]{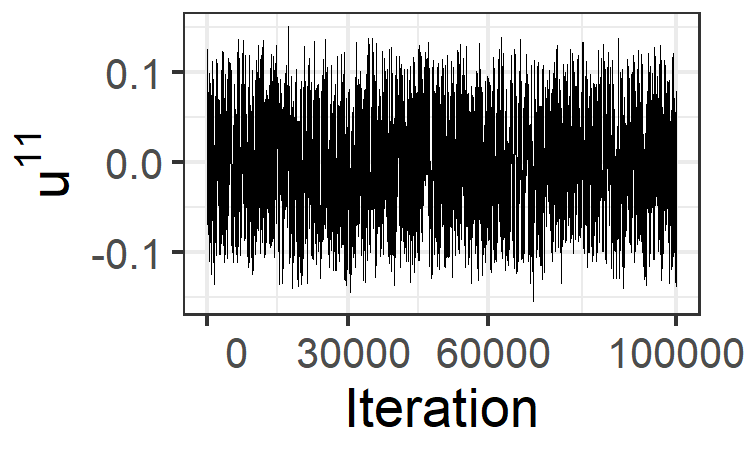}
	\includegraphics[width=.24\textwidth]{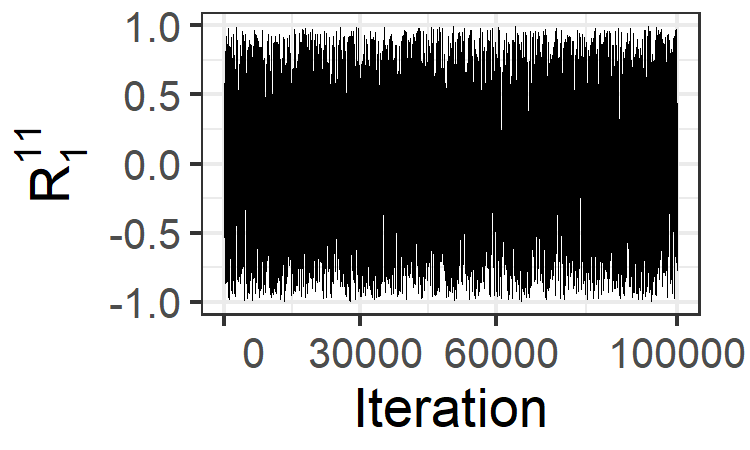}
	\includegraphics[width=.24\textwidth]{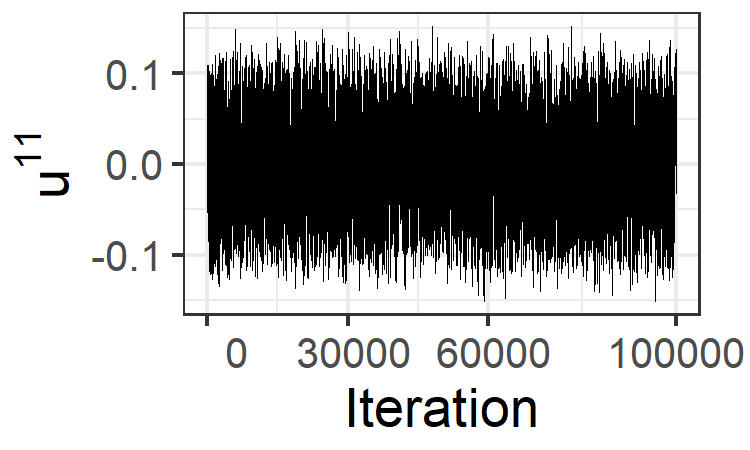}
	\caption{Traces for the Markov chains of posterior samples of the first element of parameters $R_{1}$ and $u$ from the Gibbs posterior (left two) and the gradient-bridged posterior (right two).}
	\label{fig:data_integration_trace}
\end{figure}

Figure \ref{fig:data-integration-acf} shows autocorrelation functions for posterior sampling Markov chains of all elements of rotation matrices $R_b$'s and all elements of $u$, using the gradient-bridged posterior and the Gibbs posterior.
Figure~\ref{fig:data_integration_trace} presents trace plots for the Markov chains of the posterior samples for the first element of the first rotation matrix $R_{1}$ and the first element of the shared latent parameter $u$, using two methods. The gradient-bridged posterior shows much better mixing compared to the Gibbs posterior.

\end{document}